\documentclass[sigconf]{acmart} 
\usepackage{multirow}
\usepackage{caption}
\usepackage{booktabs}
\usepackage{enumitem}
\usepackage{graphicx}
\usepackage{float} 
\AtBeginDocument{%
  }

\setcopyright{acmlicensed}
\copyrightyear{2018}
\acmYear{2018}
\acmDOI{XXXXXXX.XXXXXXX}

\acmConference[Conference acronym 'XX]{Make sure to enter the correct
  conference title from your rights confirmation emai}{June 03--05,
  2018}{Woodstock, NY}
\acmISBN{978-1-4503-XXXX-X/18/06}

\acmSubmissionID{272}

\begin{document}

\title{Towards Personalized Federated Multi-Scenario Multi-Task Recommendation}


\author{Yue Ding}
  \email{dingyue@sjtu.edu.cn}
  \affiliation{%
  \institution{Shanghai Jiao Tong University}
  \country{China}
}

\author{Yanbiao Ji}
  \email{jiyanbiao@sjtu.edu.cn}
  \affiliation{%
  \institution{Shanghai Jiao Tong University}
  \country{China}
}

\author{Xun Cai}
  \email{caixun@sjtu.edu.cn}
  \affiliation{%
  \institution{Shanghai Jiao Tong University}
  \country{China}
}

\author{Xin Xin}
  \email{xinxin@sdu.edu.cn}
  \affiliation{%
  \institution{Shan Dong University}
  \country{China}
}

\author{Yuxiang Lu}
  \email{luyuxiang\_2018@sjtu.edu.cn}
  \affiliation{%
  \institution{Shanghai Jiao Tong University}
  \country{China}
}

\author{Suizhi Huang}
  \email{huangsuizhi@sjtu.edu.cn}
  \affiliation{%
  \institution{Shanghai Jiao Tong University}
  \country{China}
}

\author{Chang Liu}
  \email{isonomialiu@sjtu.edu.cn}
  \affiliation{%
  \institution{Shanghai Jiao Tong University}
  \country{China}
}

\author{Xiaofeng Gao}
  \email{gao-xf@sjtu.edu.cn}
  \affiliation{%
  \institution{Shanghai Jiao Tong University}
  \country{China}
}

\author{Tsuyoshi Murata}
  \email{murata@c.titech.ac.jp}
  \affiliation{%
  \institution{Tokyo Institute of Technology}
  \country{Japan}
}

\author{Hongtao Lu}
  \email{htlu@sjtu.edu.cn}
  \affiliation{%
  \institution{Shanghai Jiao Tong University}
  \country{China}
}









\begin{abstract}

In modern recommender systems, especially in e-commerce, predicting multiple targets such as click-through rate (CTR) and post-view conversion rate (CTCVR) is common. Multi-task recommender systems are increasingly popular in both research and practice, as they leverage shared knowledge across diverse business scenarios to enhance performance. However, emerging real-world scenarios and data privacy concerns complicate the development of a unified multi-task recommendation model.

In this paper, we propose PF-MSMTrec, a novel framework for personalized federated multi-scenario multi-task recommendation. In this framework, each scenario is assigned to a dedicated client utilizing the Multi-gate Mixture-of-Experts (MMoE) structure. To address the unique challenges of multiple optimization conflicts, we introduce a bottom-up joint learning mechanism. First, we design a parameter template to decouple the expert network parameters, distinguishing scenario-specific parameters as shared knowledge for federated parameter aggregation. Second, we implement personalized federated learning for each expert network during a federated communication round, using three modules: federated batch normalization, conflict coordination, and personalized aggregation. Finally, we conduct an additional round of personalized federated parameter aggregation on the task tower network to obtain prediction results for multiple tasks. Extensive experiments on two public datasets demonstrate that our proposed method outperforms state-of-the-art approaches. The source code and datasets will be released as open-source for public access.

\end{abstract}

\ccsdesc[500]{Information systems~Recommender systems}

\keywords{Multi-task Recommendation, Federated Learning, Collaborative Filtering}


\received{20 February 2007}
\received[revised]{12 March 2009}
\received[accepted]{5 June 2009}

\maketitle

\section{Introduction}
Recommender systems leverage users' past behaviors to predict their interests and preferences, thereby delivering tailored recommendations for personalized content.
In modern applications of recommender systems, there are often multiple prediction targets. For example, in e-commerce scenarios, it's necessary to estimate both the click-through rate and conversion rate of products. In short video platforms, predictions may involve estimating clicks, playback time, shares, comments, and likes. Therefore, recommender systems should possess the capability to simultaneously perform multiple recommendation tasks to meet the diverse needs of users~\cite{wang2023multi}.
Traditional recommendation models usually build separate prediction models for different recommendation tasks and then merge them. Nevertheless, this model fusion approach has two main drawbacks: ($i$) Most mainstream recommendation models rely on deep neural networks with many parameters. Trying to optimize and merge multiple models simultaneously requires a lot of computational resources, making it difficult for practical online applications. ($ii$) There could be connections between different tasks, and optimizing them individually might overlook these relationships. Hence, there's a growing interest in multi-task recommender systems in both research and practical applications. 

Conventional multi-task recommender systems primarily focus on business data from a single scenario with the aim of simultaneously improving the prediction performance of multiple tasks. When there are more fine-grained businesses in the system, it is necessary to merge and model different recommendation business scenarios to utilize the commonalities between different scenarios to enhance the overall performance. Taking the example of food recommendation on Meituan~\cite{zhou2023hinet}, its business may involve various scenarios such as limited-time flash sale recommendation, search result sorting, and discounted meal package recommendation. Users engage in clicking, browsing, and other actions across multiple scenarios, eventually leading to a purchase. Such businesses can be implemented using a unified framework for multi-scenario multi-task recommendation.
\begin{figure*}[htbp]
\centering
\includegraphics[width=0.98\linewidth,height=4cm]{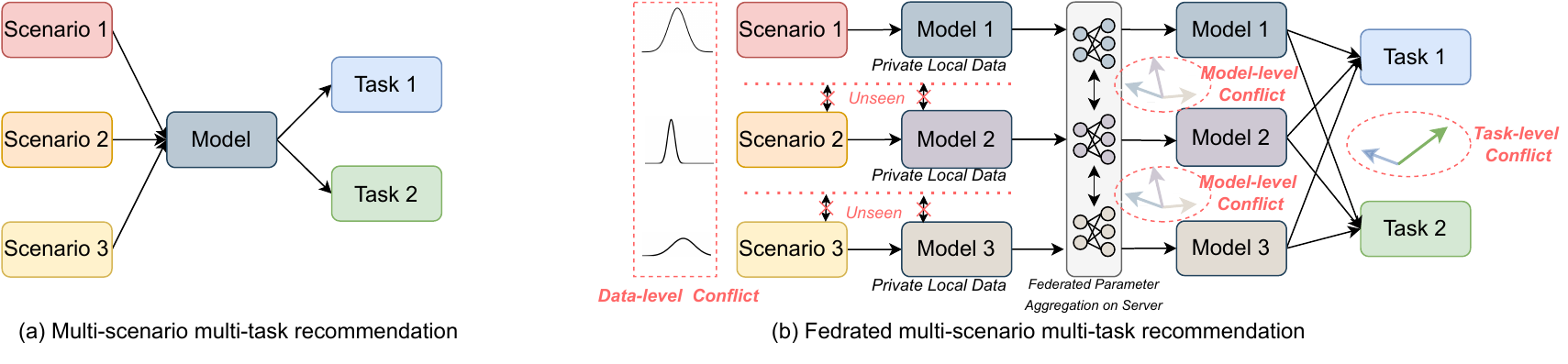}
  \caption{Illustration of non-federated (a) vs. federated multi-scenario multi-task recommendation (b). In the federated setting, diverse data distributions across scenarios (clients) cause data-level conflicts. Data and model parameters stay private on each client, while the server aggregates parameters, leading to model-level conflicts. Each client also handles multiple tasks, introducing task-level conflicts.}
\label{mtl-sample}
\end{figure*}

As real-world business is evolving rapidly, some new and more intricate recommendation business scenarios have emerged.
For example, recommender systems can leverage data from multinational corporations' global branches. This data can be segmented by user location, allowing the system to employ country-specific models that treat users from different nations as distinct scenarios. This approach accounts for potential variations in user preferences across geographical regions.
In another real-world application, advertising alliance recommendations involve a mediating platform that aggregates advertising inventory from multiple websites. This platform then connects advertisers with these websites, allowing them to display ads and earn revenue based on factors like ad impressions or clicks.
In these two cases, privacy concerns arise because data from each participant is private, and individual prediction models are customized. Consequently, the issue is that training a single, global model becomes very difficult.

Federated learning (FL)~\cite{konevcny2016federated} is a collaborative learning paradigm that allows joint optimization across multiple clients while preserving all clients' data privacy. 
However, it is difficult to simply extend multi-scenario and multi-task recommendation to the FL framework. The key challenging issue is that multiple optimization conflicts overlap and intertwine in this situation, which can easily lead to a decline in overall performance. 
More specifically, ($i$) \emph{\textbf{Data-level conflict}.} The data distributions of user-item interactions in multiple scenarios vary. In FL, each client has an independent model to fit the scenario-specific data, and all the clients' models may project data to different feature spaces because each client's data is private and isolated. ($ii$) \emph{\textbf{Model-level conflict}.} In FL, models from different clients are typically aggregated using parameter averaging. Differences in data distribution among clients can lead to discrepancies in model parameters, resulting in performance decline on all clients during federated parameter aggregation. ($iii$) \emph{\textbf{Task-level conflict}.} Different tasks have different targets which may influence each other. If the model cannot effectively balance these interdependent targets, it can prioritize one task over another, leading to uneven performance, which is recognized as the "task seesaw phenomenon"~\cite{tang2020progressive}. 
Figure~\ref{mtl-sample} illustrates the difference and federated multi-scenario multi-task recommendation and non-federated scenario.
To the best of our knowledge, federated learning has not yet been explored for multi-scenario multi-task recommendation.

To address the research gap and tackle this challenging problem, we propose a \textbf{P}ersonalized \textbf{F}ederated learning framework for \textbf{M}ulti-\textbf{S}cenario \textbf{M}ulti-\textbf{T}ask recommendation (PF-MSMTrec). 
Specifically, we assign each scenario to a distinct client, ensuring that the training data for each client remains independent and private. We utilize a Multi-gate Mixture-of-Experts (MMoE) structure for each client, where each client may have multiple expert networks. We decouple the parameters of each expert network into three categories: global shared parameters, task-specific parameters, and scenario-specific parameters, by designing a parameter template. Federated aggregation is then performed on the scenario-specific parameters, while the other parameters are treated as local personalized parameters. We propose a set of modules for federated parameter aggregation, including federated batch normalization, conflict coordination, and personalized aggregation, all of which are utilized during the federated communication rounds. Finally, after processing through the local gate network, we apply the conflict coordination mechanism to the parameters of the tower networks for personalized aggregation, facilitating multi-task prediction.

The main contributions of this paper are summarized as follows:
\begin{itemize}[leftmargin=*]
\item We propose a novel personalized federated recommendation framework for multi-scenario multi-task recommendation. To the best of our knowledge, it is the first work to tackle this challenging problem. The proposed method broadens the applicability of recommender systems by tackling more sophisticated business settings.
\item To address the multiple optimization conflicts inherent in federated multi-scenario multi-task recommendation, we propose a meticulously designed bottom-up joint learning mechanism, which incorporates modules for expert network parameter decoupling, federated batch normalization, conflict coordination, and personalized parameter aggregation. The collaborative work of these modules effectively alleviate optimization conflicts and enable personalized learning for local models.
\item We conduct extensive experiments on two public datasets, thoroughly comparing the performance of the proposed method with state-of-the-art (SOTA) multi-scenario multi-task recommendation methods and federated learning approaches. Notably, by adapting our method to the federated learning setting, we exceed the performance of SOTA methods originally designed for traditional centralized learning.
\end{itemize}

\section{Related Work}
\subsection{Multi-task and Multi-scenario Recommendation}
Multi-task learning has been widely researched and applied in the fields of Computer Vision (CV)~\cite{mtl-cv1} and Natural Language Processing (NLP)~\cite{mtl-nlp1}. In the area of recommender systems, multi-task recommendation is generally based on deep learning and mainly includes three categories of techniques~\cite{wang2023multi}: Optimization methods, Training mechanisms, and Parameter sharing. Optimization methods refer to addressing the problem of performance degradation due to gradient conflicts among multiple tasks during parameter training~\cite{he2022metabalance}. Training mechanism techniques are aimed at setting training and learning strategies for different tasks~\cite{bai2022contrastive}. Parameter sharing methods are the most important category among them.
The Multi-gate Mixture-of-Expert (MMoE)~\cite{ma2018modeling} model is the mainstream architecture for multi-task recommender systems. MMoE is a parameter sharing approach at the expert network level, which combines multiple expert networks using different gates to make predictions for multiple tasks. PLE ~\cite{tang2020progressive} designs shared experts and task-exclusive experts, customizing learning for different tasks using a customized gating module. ESMM~\cite{ma2018entire} implicitly trains the target task using auxiliary tasks. AITM~\cite{xi2021modeling} introduces sequence dependency based on ESMM, enhancing the modeling of inter-task correlations by introducing a self-attention sequence dependency propagation module. 
CSRec~\cite{bai2022contrastive} is a contrastive learning-based method that can adaptively update parameters to alleviate task conflicts. 
CMoIE~\cite{wang2022multi} improves the mixture policy for multiple expert networks by devising conflict resolution, expert communication, and mixture calibration modules. 
AdaTT~\cite{li2023adatt} uses an adaptive fusion mechanism to jointly learn task-specific and shared features.  

Multi-scenario recommendation systems enhance understanding of user behaviors and preferences across different domains to provide more personalized recommendations.
STAR~\cite{sheng2021one} decouples domain parameters and uses a star topology for multi-domain recommendation.
AFT~\cite{hao2021adversarial} employs generative adversarial networks for feature translation between domains.
HAMUR~\cite{li2023hamur} uses adapter layers for multi-domain recommendation.
EDDA~\cite{ning2023multi} disentangles embeddings and aligns domains to improve knowledge transfer.
MetaDomain~\cite{zhang2023meta} uses a domain intent extractor and meta-generator to fuse domain intent representations for predictions.
ADIN~\cite{jiang2022adaptive} models commonalities and differences across scenarios with an adaptive domain interest network.
Maria~\cite{tian2023multi} injects scene semantics at the network's bottom for adaptive feature learning.
PLATE~\cite{wang2023plate} proposes a pre-train and prompt-tuning paradigm to efficiently enhance performance for multiple scenarios. 
SAMD~\cite{huan2023samd} addresses the multi-scenario heterogeneity problem through knowledge distillation.

Modeling multiple scenarios and tasks simultaneously is currently a topic of growing interest in the field of recommender systems.
AESM$^2$~\cite{zou2022automatic} proposes an automatic expert search framework for multi-task learning, integrating hierarchical multiple expert layers with different recommendation scenarios.
PEPNet~\cite{chang2023pepnet} is a plug-and-play parameter and embedding personalized network suitable for multi-scenario and multi-task recommendations.
HiNet~\cite{zhou2023hinet} is a multi-scenario multi-task recommendation model based on a hierarchical information extraction network. It achieves information extraction through a knowledge transfer scheme from coarse-grained to fine-grained levels.
M3REC~\cite{lan2023m3rec} is a meta-learning-based framework that realizes unified representations and optimization in multiple scenarios and tasks.

\subsection{Federated Learning for Recommendation}
Federated learning is a distributed machine learning framework for preserving data privacy, mainly using the method of passing model parameters to implicitly coordinate the training of models among various participants. According to the differences in data and feature dimensions among different participants, federated learning can be roughly divided into three categories: horizontal federated learning, vertical federated learning, and federated transfer learning~\cite{yang2019federated}.
Typical federated learning methods include:
FedAvg~\cite{mcmahan2017communication}: It calculates the average of model parameters from all participants as the global model parameters.
Personalized Federated Learning (PFL) aims to alleviate the slow convergence and poor performance problems under non-i.i.d. (non-independent and identically distributed) data, making the model personalized for local tasks and datasets~\cite{tan2022towards}.

Federated recommender systems are one of the important applications of federated learning. 
FedRecSys~\cite{tan2020federated} is an open-source federated recommendation systems capable of providing online services. 
Many collaborative filtering algorithms also have corresponding federated learning versions, such as Federated Collaborative Filtering (FCF)~\cite{ammad2019federated} and Federated Matrix Factorization~\cite{lin2020fedrec}.
FedFast~\cite{muhammad2020fedfast} and FL-MV-DSSM~\cite{huang2020federated} are representative federated recommender systems based on deep learning techniques.
DeepRec~\cite{han2021deeprec} proposes federated sequence recommendation.  FedGNN~\cite{wu2021fedgnn}, PerFedRec~\cite{luo2022personalized}, and SemiDFEGL~\cite{qu2023semi} are graph neural network-based federated recommendation models. 
PFedRec~\cite{zhang2023dual} is a cross-device personalized federated recommendation framework that learns lightweight models to capture fine-grained user and item features. 
RF$^2$~\cite{maeng2022towards} investigates the fairness problem in federated recommendation, and F$^2$PGNN~\cite{agrawal2024no} further addresses the group bias issue in graph neural networks, proposing a fair and personalized federated recommendation framework.   

\subsection{Federated Multi-task Learning}
Federated multi-task learning has emerged as a research problem in recent years~\cite{he2022spreadgnn,mills2021multi}. 
Approaches such as MOCHA~\cite{smith2017federated} and FedEM~\cite{marfoq2021federated} have proposed methods for jointly training multiple tasks across multiple participants with diverse data distributions.
Fedbone~\cite{chen2023fedbone} enhances feature extraction capability by aggregating encoders from gradients uploaded by each client.
Addressing the issue of task heterogeneity, MAS~\cite{zhuang2023mas} allocates different multi-task models to different clients and aggregates models within clients with the same task set.
MaT-FL~\cite{cai2023many} uses dynamic grouping to combine different client models.

\section{Method}
\begin{figure*}[htbp]
\centering
\includegraphics[width=0.93\linewidth]{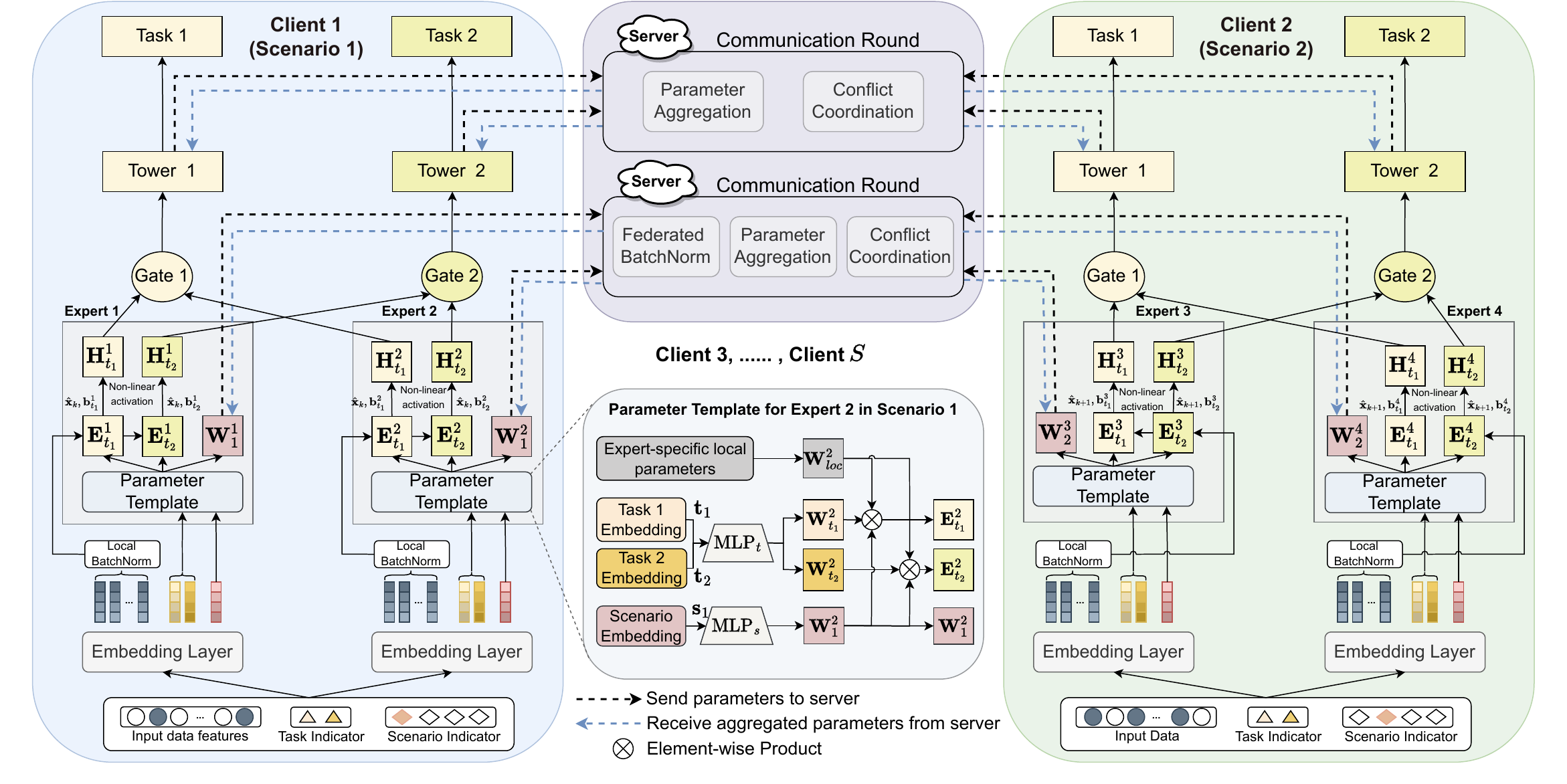}
  \caption{Framework of PF-MSMTrec for personalized federated multi-scenario multi-task recommendation. Each client handles a unique scenario with private data, using the MMoE structure. Parameter decoupling in expert networks enables federated aggregation of scenario-specific features. Federated batch normalization, conflict coordination, and personalized aggregation are applied in each communication round to address optimization conflicts. For clarity, Expert 3 and Expert 4 are used for Client 2, with $\hat{\mathbf{x}}_k$ and $\hat{\mathbf{x}}_{k+1}$ representing different inputs.}
\label{architecture}
\end{figure*}

Our proposed PF-MSMTrec method is illustrated in Figure~\ref{architecture}. We detail our framework components below.
\subsection{Problem Definition}
Suppose we have a total of $S$ application scenarios, each separated from the others. 
Correspondingly, we have $S$ clients,  with each client restricted to accessing only local data. 
We employ the mixture-of-experts structure, allowing each client to utilize multiple expert networks to handle various prediction tasks.
Assuming there are a total of $T$ tasks and $S$ clients, where each client comprises $N$ experts.   
We define the problem within the $j$-th client for predicting the $i$-th task as follows:
\begin{equation}
    \hat{y}_{i} = \text{Fed} \ [\parallel_{n=1}^{N} f_{i}^n (\mathbf{x}, t_i, s_j | D_{j})], 
\end{equation}
where $i \in \{1,...,T\}$ and $j \in \{1,...,S\}$. $\text{Fed} \ [\parallel_{n=1}^{N} f_{i}^n (\cdot)]$ represents the prediction of the $i$-th task, which is made by a combination of $N$ expert networks within the context of federated learning. Here, $\mathbf{x}$ denotes the dense feature, $D_{j}$ represents local data in the $j$-th scenario, $t_i$ and $s_j$ denote the task and scenario embeddings, respectively.

\subsection{Decoupling Expert Parameters}
For the $j$-th client that has $N$ experts, we input three types of data into each expert network: dense feature vectors, task feature vectors, and scenario feature vectors.
For dense feature vectors, we apply Batch Normalization (BN):
\begin{equation}
  \mathbf{\mu} = \frac{1}{K} \sum_{k=1}^{K} \mathbf{x}_{k}, \ \ 
  \mathbf{\sigma}^2 = \frac{1}{K} \sum_{k=1}^{K} (\mathbf{x}_{k}-\mu)^2, \ \   \hat{\mathbf{x}}_k = \gamma \cdot \frac{\mathbf{x}_k -\mathbf{\mu}}{\sqrt{\mathbf{\sigma}^2+\mathbf{\epsilon}}} + \mathbf{\beta},
\label{domainbn-1}
\end{equation}
where $K$ represents the number of data samples in one batch, $\mathbf{x}_k$ denotes the $k$-th input dense feature vector, $\gamma$ and $
\beta$ are learnable parameters, and $\epsilon$ is a small constant. 
BN normalizes local input data and mitigates discrepancies in data distribution at the input layer, thereby alleviating conflicts arising from data distribution.

To generate task-specific parameters, we employ a parameter template comprising neural networks dedicated to each expert.
This design facilitates parameter separation within expert networks, enabling them to extract not only input features but also task-related and scenario-specific features:
\begin{equation}
   \mathbf{W}_{t_i}^n = \text{MLP}_t(\mathbf{t}_i), \  \  \mathbf{W}_{j}^n = \text{MLP}_s(\mathbf{s}_j), 
\end{equation}
where $n \in \{1,2,..., N\}$ indicates the number of the expert network, $\text{MLP}_t$, and $\text{MLP}_s$ represent multi-layer perceptrons capable of generating parameters based on the input. $\mathbf{t}_i$ and $\mathbf{s}_j$ are the task and scenario dense embeddings, respectively. 
By introducing the expert-specific local parameter $\mathbf{W}_{loc}^n$, the local task-specific parameters are composed of the following three components: 
\begin{equation}
\mathbf{E}_{t_i}^n=\mathbf{W}_{loc}^n\otimes\mathbf{W}_{t_i}^n\otimes\mathbf{W}_{j}^n,
\label{exp_task_para}
\end{equation}
where $\otimes$ denotes element-wise product. Subsequently, task-specific feature outputs can be obtained based on the input embedding:
\begin{equation}
  \mathbf{H}_{t_i}^n = \sigma(\hat{\mathbf{x}}_k\cdot\mathbf{E}_{t_i}^n +\mathbf{b}_{t_i}^{n}),
  \label{exp_out}
\end{equation}
where $\sigma$ represents the ReLU non-linear activation function, and $\mathbf{b}_{t_i}^n$ is the bias term. 
For all experts within the $j$-th client, the output set for the $i$-th task is described as the following set:
\begin{equation}
\{\text{Exp}_i\}_j = 
\{ \mathbf{H}_{t_i}^1, \mathbf{H}_{t_i}^2, ..., \mathbf{H}_{t_i}^N 
\}_j.
\label{exp_out_all}
\end{equation}
Here we use $\{\cdot\}_j$ to denote the set of outputs for the $j$-th client to avoid confusion caused by using superscripts and subscripts. We also emphasize that the parameters $\mathbf{W}_{loc}^n$, $\mathbf{W}_{t_i}^n$, and $\mathbf{W}_{j}^n$ are distinct for each client.

\subsection{Federated Parameter Aggregation}
\subsubsection{Fedrated Batch Normalization}
Under the federated learning paradigm, we jointly train expert and tower networks using parameter aggregation. 
To implement personalization, we designate $\mathbf{W}_{loc}^n$ and $\mathbf{W}_{t_i}^n$ as local parameters, and $\mathbf{W}_{j}^n$ as shared parameter for federated learning.
We designed this approach to let each expert understand different tasks in various scenarios. We avoid federated aggregation on task parameters due to task commonalities and instead share the scenario-specific parameter $\mathbf{W}_{j}^n$ among clients to facilitate mutual learning.
Additionally, as shown in Eq.~(\ref{exp_task_para}), $\mathbf{W}_{j}^n$ is integrated into the local parameters, enabling updates to $\mathbf{W}_{j}^n$ to influence the learning of personalized local parameters.
We employ a simple federated batch normalization strategy, treating all shared expert network parameters as a batch on the server during each communication round.
\begin{equation}
\mathbf{\mu}_{g} = \frac{1}{N\cdot S} \sum_{n=1}^{N} \sum_{j=1}^{S}\mathbf{W}_{j}^n, \ \mathbf{\sigma}_{g}^2 = \frac{1}{N\cdot S} \sum_{n=1}^{N} \sum_{j=1}^{S} (\mathbf{W}_{j}^n - \mathbf{\mu}_{g})^2,
\label{fedbn:u}
\end{equation}
\begin{equation}
\text{FedBN}(\mathbf{W}_{j}^n) = \gamma_{g} \cdot  \frac{\mathbf{W}_{j}^n -\mathbf{\mu}_{g}}{\sqrt{\mathbf{\sigma}_{g}^2+\mathbf{\epsilon}_g}} + \mathbf{\beta}_{g},
\label{fedbn:bn}
\end{equation}
where the subscript $g$ represents global. It is important to note that we calculate $\mathbf{\gamma}_g$ and $\beta_g$ by averaging the local $\mathbf{\gamma}$ and $\mathbf{\beta}$ values from all clients, as there are no learnable parameters on the server.
\begin{equation}
\mathbf{\gamma}_{g}, \mathbf{\beta}_g = \frac{1}{S} \sum_{j=1}^{S} \mathbf{\gamma}_j,\mathbf{\beta}_j.
\end{equation}
The typical federated aggregation approach is parameter averaging:
\begin{equation}
\bar{\mathbf{W}}_{j}^n
= \frac{1}{N\cdot S} \sum_{n=1}^{N} \sum_{j=1}^{S}
\text{FedBN}(\mathbf{W}_{j}^n).
\label{fed_ws_avg}
\end{equation}
Interestingly, we found that $\bar{\mathbf{W}}_{j}^n$ is only related to normalization parameter $\mathbf{\beta}_g$ if federated communication occurs after the completion of each local batch. We provide a proof in Appendix A. It's important to note that local normalization operates on data, while federated normalization operates on parameters uploaded by clients. Their unification shows that parameter normalization during federated aggregation effectively leverages client data characteristics, addressing optimization conflicts due to domain distribution differences. 
\subsubsection{Conflict Coordination}
Large parameter differences between clients during federated parameter aggregation are the cause of performance degradation in federated learning. Since parameters are determined by the gradients from local training, alleviating gradient conflict would be a direct and effective approach.
Considering the local shared parameter $\mathbf{W}_{j}^n$ in the expert network, we update:
\begin{equation}
\mathbf{W}_{j}^{n'} \leftarrow \mathbf{W}_{j}^n - \eta \sum_{i=1}^T \mathbf{g}_i,
\end{equation}
where $\eta$ is the learning rate, $\mathbf{g}_i$ denotes the gradient on the $i$-th task loss. We define the increment of the parameter update as $\Delta \mathbf{W}_{j}^n = - \eta \sum_{i=1}^T \mathbf{g}_i$. 
Similarly, we have the global average increment of aggregated parameters:
\begin{equation}
\Delta \bar{\mathbf{W}}_{j}^n = - \eta \frac{1}{N\cdot S} \sum_{n=1}^{N} \sum_{j=1}^{S}\sum_{i=1}^T \{\mathbf{g}_i^n \}_j.
\end{equation}
Drawing inspiration from CAGrad~\cite{liu2021conflict}, we identify a set of parameters, denoted as $\mathbf{U}$, to mitigate gradient conflicts between $\Delta \mathbf{W}_{j}^n$ and $\Delta \bar{\mathbf{W}}_{j}^n$ through a simple gradient inner product and gradient constraint. This is achieved by optimizing the following objective:
\begin{equation}
    \max_{\mathbf{U}} \ \min_{} \langle \Delta \mathbf{W}_{j}^n,\mathbf{U}\rangle \quad \mathrm{s.t.} \|\mathbf{U}-\Delta \bar{\mathbf{W}}_{j}^n\|\leq c \|\Delta \bar{\mathbf{W}}_{j}^n\|,
    \label{fed_cagrad}
\end{equation}
where $\langle \cdot, \cdot \rangle$ denote inner product, $c \in [0,1)$ is the hyper-parameter.
Note that federated learning generally restricts clients from uploading gradients due to privacy and security concerns. Therefore, to approximate the gradient, we calculate the difference between the parameters from two consecutive communication rounds:
\begin{equation}
    \Delta \hat{\mathbf{W}}_j^n = [\bar{\mathbf{W}}_j^n]^{r} - [\bar{\mathbf{W}}_j^n]^{r-1},
\end{equation}
where $r$ denotes the communication round. Substitute $\Delta \hat{\mathbf{W}}_{n,j}$ for $\Delta \mathbf{W}_{n,j}$ in Eq.~(\ref{fed_cagrad}), we can solve the optimization problem by using Lagrangian and adding the constraint of $\sum_{n=1}^{N} \sum_{j=1}^{S} w_{n,j}=1, w_{n,j}\ge 0$, we turn to the following optimization problem for $w$:
\begin{align}
\label{cagrad}
    &\min_w F(w)=\mathbf{U}_w^\top \cdot \Delta \hat{\mathbf{W}}_j^n + \sqrt{\phi} \cdot \|\mathbf{U}_w\|, \\
    &\text{where} \ \mathbf{U}_w= \sum_{n=1}^{N} \sum_{j=1}^{S} w_{n,j} \cdot \Delta \hat{\mathbf{W}}_j^n, \ \text{and} \ \phi=c^2\|\Delta \hat{\mathbf{W}}_j^n\|^2.
\end{align}
We can derive the solution of $\mathbf{U}^*$ after obtaining the optimal $w$:
\begin{equation}
\mathbf{U}^* 
 = \Delta \hat{\mathbf{W}}_j^n + \frac{\sqrt{\phi}}{\|U_w\|}U_w. \label{eq:enc_final}
\end{equation}

\subsubsection{Personalized Parameter Aggregation}
To maintain the personalization of local parameters while effectively aggregating federated parameters, we introduce a learnable weight parameter, $\psi_n$ for each expert network. The updating process of the shared parameter $\mathbf{W}_{j}^n$ is represented as:
\begin{equation}
[\mathbf{W}_{j}^n]^{r}=[\mathbf{W}_{j}^n]^{r-1}+[\Delta \bar{\mathbf{W}}_{j}^n]^{r}+\psi_n \mathbf{U}^*,
\end{equation}
For the gate network that is responsible for aggregating outputs from various experts for the $i$-th task in the $j$-th scenario, we have:
\begin{equation}
    \text{Gate} [\{\text{Exp}_i\}_j ] = (a_1 \mathbf{H}_{t_i}^1 + a_2 \mathbf{H}_{t_i}^2+,...,+a_N \mathbf{H}_{t_i}^N)_j,
\end{equation}
where $a_1,...,a_N$ are weight parameters learned by a neural network, subject to $a_n \geq 0$ and $\sum_{n=1}^N a_n= 1$.
The task tower, responsible for generating predictions for the $i$-th task, is defined as follows:
\begin{equation}
    \hat{p}_{i,j} = \text{MLP} (\text{Gate}[\{\text{Exp}_i\}_j] ).
\end{equation}
We conduct a personalized federated aggregation of all parameters within the Tower network.
\begin{equation}
[\Theta_{i,j}^{\text{Tow}}]^{r}=[\Theta_{i,j}^{\text{Tow}}]^{r-1}+[\Delta\Theta_{i,j}^{\text{Tow}}]^{r}+\psi_n' \mathbf{U}^*,
\end{equation}
where $\Theta_{i,j}^{\text{Tow}}$ denotes the parameters of the $i$-th task tower in the $j$-th scenario, $\psi_n'$ represents another learnable weight parameter.

\subsection{Optimization}
We utilize the binary cross-entropy loss as the loss function:
\begin{equation}
  \mathcal{L}_{BCE_j} = \frac{1}{|\mathcal{X}_j|} \sum_{\mathbf{x}=1}^{|\mathcal{X}_j|} \sum_{i=1}^{T} -\mathbf{y} \text{log}(\mathbf{\hat{p}}_{i,j}) - (1-\mathbf{y})\text{log}(1-\mathbf{\hat{p}}_{i,j}),
\label{bce}
\end{equation}
where $|\mathcal{X}|_j$ represents the number of data samples in $D_j$. To align local shared parameters with global aggregated parameters and prevent performance degradation, we introduce an additional regularization term:
\begin{equation}
    \mathcal{L}_{d_j} = \sum_{n=1}^{N} \Vert \bar{\mathbf{W}}_j^n -\mathbf{W}_j^n \Vert_{2}^{2}.
\end{equation}
The total loss for the $j$-th scenario is:
\begin{equation}
    \mathcal{L}_j = \mathcal{L}_{BCE_j} + \lambda \mathcal{L}_{d_j},
\end{equation}
where $\lambda$ is a manually set coefficient. 

\section{Experiments}
\subsection{Experimental Settings}
\subsubsection{Datasets}
We conduct experiments on two public datasets: 
(1) \textbf{AliExpress Dataset\footnote{\url{https://tianchi.aliyun.com/dataset/74690}}}. It is collected from a real-world search system in AliExpress, we use four scenarios: Netherlands (NL), Spain (ES), France (FR), and the United States (US). 
(2) \textbf{Tenrec Benchmark\footnote{\url{https://github.com/yuangh-x/2022-NIPS-Tenrec}}}. Tenrec is a dataset suite for multiple recommendation tasks, collected from two different recommendation platforms of Tencent, QQ BOW (QB) and QQ KAN (QK). Items in QK/QB can be news articles or videos. We use data from two scenarios, QK-video and QB-video. Table~\ref{tab:stat} describes the statistics of the datasets.
\begin{table}[H]
\small
  \caption{Statistics of test datasets.}
  \label{tab:stat}
  \begin{tabular}{ccccc}
    \toprule
    \textbf{Dataset} &\textbf{Train} &\textbf{Validation} &\textbf{Test} &\textbf{\#Features}\\
    \midrule
    AliExpress NL & 12.1M & 5.6M & 5.6M & 79\\
    AliExpress ES & 22.3M & 9.3M & 9.3M & 79\\
    AliExpress FR & 18.2M & 8.8M & 8.8M & 79\\
    AliExpress US & 19.9M & 7.5M & 7.5M & 79\\
    Tenrec-QK-video & 69.3M & 8.7M & 8.7M & 16\\
    Tenrec-QB-video & 1.9M & 0.2M & 0.2M & 16\\
  \bottomrule
\end{tabular}
\end{table}

\begin{table*}[t]
\small
  \caption{Performance comparison of multi-scenario and multi-task methods on the AliExpress dataset. Bold text indicates the overall best performance, while underlined text denotes the best performance among baseline methods. * indicates the improvements over the best baseline are statistically significant (i.e., one-sample t-test with \emph{p} < 0.05).}
  \centering
    \begin{tabular}{@{}lcccccccc@{}}
    \toprule
    \textbf{AliExpress} & \multicolumn{2}{c}{\textbf{NL}} & \multicolumn{2}{c}{\textbf{ES}} & \multicolumn{2}{c}{\textbf{FR}} & \multicolumn{2}{c@{}}{\textbf{US}} \\
    \cmidrule(lr){2-3} \cmidrule(lr){4-5} \cmidrule(lr){6-7} \cmidrule(lr){8-9}
     & auc\_ctr & auc\_ctcvr & auc\_ctr & auc\_ctcvr & auc\_ctr & auc\_ctcvr & auc\_ctr & auc\_ctcvr \\
    \midrule
    Single-task & 0.7203 & 0.8556 & 0.7252 & 0.8832 & 0.7174 & 0.8702 & 0.7058 & 0.8637 \\
    MMoE & 0.7195 & 0.8574 & 0.7269 & 0.8899 & 0.7226 & 0.8748 & 0.7053 & 0.8639 \\
    PLE & 0.7268 & 0.8571 & 0.7268 & 0.8861 & 0.7252 & 0.8679 & 0.7092 & 0.8699 \\
    ESMM & 0.7202 & 0.8606 & 0.7263 & 0.8891 & 0.7222 & 0.8684 & 0.7035 & 0.8712 \\
    AITM & 0.7256 & 0.8586 & 0.7270 & 0.8829 & 0.7216 & 0.8710 & 0.7019 & 0.8655 \\
    STAR & 0.7263 & 0.8624 & 0.7281 & 0.8891 & 0.7269 & 0.8803 & 0.7088 & 0.8765 \\
    AESM\textsuperscript{2} & 0.7260 & 0.8638 & 0.7295 & \textbf{\underline{0.8949}} & 0.7241 & 0.8808 & 0.7088 & 0.8774 \\
    PEPNet & \underline{0.7310} & \underline{0.8687} & \underline{0.7342} & 0.8915 & \underline{0.7296} & \underline{0.8813} & \underline{0.7105} & \textbf{\underline{0.8851}} \\
    \textbf{PF-MSMTrec} (Local) & \textbf{0.7330*} & \textbf{0.8690*} & \textbf{0.7364*} & 0.8927 & \textbf{0.7320*} & \textbf{0.8817*} & \textbf{0.7150*} & 0.8808 \\
   \textbf{PF-MSMTrec} (Fed) & \textbf{0.7316*} & 0.8653 & 0.7325 & 0.8925 & \textbf{0.7321*} & \textbf{0.8851*} & \textbf{0.7142*} & 0.8791 \\
    \bottomrule
    \end{tabular}%
  \label{tab:overall-ali-1}%
\end{table*}%

\begin{table*}[t]
\small
  \caption{Performance comparison with federated learning methods on the Tenrec dataset. Bold text highlights the overall best performance, while underlined text indicates the best among baseline methods. * indicates the improvements over the best baseline are statistically significant (i.e., one-sample t-test with \emph{p} < 0.05).}
  \centering
    \begin{tabular}{@{}lcccccccc@{}}
    \toprule
    \textbf{AliExpress} & \multicolumn{2}{c}{\textbf{NL}} & \multicolumn{2}{c}{\textbf{ES}} & \multicolumn{2}{c}{\textbf{FR}} & \multicolumn{2}{c@{}}{\textbf{US}} \\
    \cmidrule(lr){2-3} \cmidrule(lr){4-5} \cmidrule(lr){6-7} \cmidrule(lr){8-9}
     & auc\_ctr & auc\_ctcvr & auc\_ctr & auc\_ctcvr & auc\_ctr & auc\_ctcvr & auc\_ctr & auc\_ctcvr \\
    \midrule
    Single-task & 0.7203 & \underline{0.8556} & 0.7252 & 0.8832 & 0.7174 & 0.8702 & 0.7058 & 0.8637 \\
    FedAvg & 0.7265 & 0.8551 & 0.7280 & 0.8886 & 0.7265 & 0.8664 & 0.7084 & 0.8701 \\
    FedProx & 0.7269 & 0.8547 & 0.7281 & 0.8902 & 0.7260 & 0.8705 & 0.7088 & 0.8665 \\
    Ditto & \underline{0.7273} & 0.8549 & \underline{0.7285} & \underline{0.8906} & 0.7264 & \underline{0.8708} & \underline{0.7089} & 0.8666 \\
    FedAMP & 0.7270 & 0.8552 & 0.7282 & 0.8905 & \underline{0.7266} & \underline{0.8708} & \underline{0.7089} & 0.8665 \\
    \textbf{PF-MSMTrec} (Fed) & \textbf{0.7316*} & \textbf{0.8653*} & \textbf{0.7325*} & \textbf{0.8925*} & \textbf{0.7321*} & \textbf{0.8851*} & \textbf{0.7142*} & \textbf{0.8791*} \\
    \bottomrule
    \end{tabular}%
  \label{tab:oveall-ali-2}%
\end{table*}%

\subsubsection{Baseline Methods}
We use two different groups of baseline methods, the first group being SOTA multi-scenario and multi-task recommendation methods:
\textbf{Single-task Model}. 
It employs an MLP to predict the output for a single task.
Different tasks are optimized separately in the single-task model. 
\textbf{MMoE}~\cite{ma2018modeling}, \textbf{PLE}~\cite{tang2020progressive}, and
\textbf{ESMM}~\cite{tang2020progressive} are representative multi-task recommendation baselines.
\textbf{AITM}~\cite{xi2021modeling} models task dependencies through an attention mechanism.
Note that the two prediction tasks in Tenrec (click and like) are unrelated, we did not implement AITM on the Tenrec dataset.
\textbf{STAR}~\cite{sheng2021one} is a multi-domain model that includes both center parameters and domain-specific parameters. 
\textbf{AESM}$^2$~\cite{zou2022automatic} is an advanced multi-scenario and multi-task recommendation model with automatic expert selection.
\textbf{PEPNet}~\cite{chang2023pepnet} is the SOTA method for multi-scenario multi-task recommendation. The second group consists of SOTA federated learning methods.
\textbf{FedAvg}~\cite{mcmahan2017communication} averages the parameters of all clients in federated learning and then shares them back to each client.
\textbf{FedProx}~\cite{fedprox} tackles heterogeneity in federated learning, and can be seen as an extension of FedAvg.
\textbf{Ditto}~\cite{ditto} emphasizes fairness and robustness in personalized federated learning.
\textbf{FedAMP}~\cite{fedamp} improves collaboration among clients with non-i.i.d. data distributions through attentive message passing.
\subsubsection{Evaluation Metric}
For the AliExpress dataset, all methods perform two tasks: predicting click-through rate (CTR) and post-view click-through and conversion rate (CTCVR). For the Tenrec dataset, the tasks involve predicting user clicks and likes on exposed videos or articles.
We adopt the widely used Area Under Curve (AUC) as the evaluation metric in our experiments:
\begin{equation}
  AUC = \frac{1}{|D_{test}^{+}||D_{test}^{-}|}\sum_{x^{+} \in D_{test}^{+}}\sum_{x^{-} \in D_{test}^{-}}I(f(x^{+}) > f(x^{-})),
\end{equation}
where $D_{test}^{+}$ and $D_{test}^{-}$ represent the collections of positive and negative samples in the test set, respectively. $f(\cdot)$ denotes the prediction function, and $I(\cdot)$ denotes the indicator function.

\begin{table}[t]
\small
  \caption{Performance comparison of multi-scenario and multi-task methods on the Tenrec dataset. Bold text highlights the overall best performance, while underlined text denotes the best performance among baseline methods. * indicates the improvements over the best baseline are statistically significant (i.e., one-sample t-test with \emph{p} < 0.05).}
  \centering
    \begin{tabular}{@{}lcccccccc@{}}
    \toprule
    \textbf{Tenrec} & \multicolumn{2}{c}{\textbf{QK-video}} & \multicolumn{2}{c}{\textbf{QB-article}} \\
    \cmidrule(lr){2-3} \cmidrule(lr){4-5}
     & auc\_click & auc\_like & auc\_click & auc\_like \\
    \midrule
    Single-task & 0.7957 & 0.9160 & 0.8013 & 0.9343 \\
    MMoE & 0.7900 & 0.9020 & 0.8002 & 0.9212 \\
    PLE & 0.7822 & 0.9103 & 0.8031 & 0.9310 \\
    ESMM & 0.7898 & 0.9089 & 0.8024 & 0.9285 \\
    STAR & 0.7920 & 0.9188 & 0.8055 & 0.9314 \\
    AESM\textsuperscript{2} & 0.7942 & \underline{0.9219} & 0.8047 & 0.9297 \\
    PEPNet & \underline{0.7953} & 0.9200 & \underline{0.8076} & \textbf{\underline{0.9331}} \\
    \textbf{PF-MSMTrec} (Local) & \textbf{0.7956*} & \textbf{0.9221*} & \textbf{0.8080*} & 0.9321 \\
    \textbf{PF-MSMTrec} (Fed) & \textbf{0.7965*} & 0.9202 & \textbf{0.8083*} & 0.9327 \\
    \bottomrule
    \end{tabular}%
  \label{tab:overall-tenrec-1}%
\end{table}%

\begin{table}[t]
\small
  \caption{Performance comparison of federated learning methods on the Tenrec dataset. Bold text indicates the overall best performance, while underlined text denotes the best performance among baseline methods. * indicates the improvements over the best baseline are statistically significant (i.e., one-sample t-test with \emph{p} < 0.05).}
  \centering
    \begin{tabular}{@{}lcccccccc@{}}
    \toprule
    \textbf{Tenrec} & \multicolumn{2}{c}{\textbf{QK-video}} & \multicolumn{2}{c}{\textbf{QB-article}} \\
    \cmidrule(lr){2-3} \cmidrule(lr){4-5}
     & auc\_click & auc\_like & auc\_click & auc\_like \\
    \midrule
    Single-task & 0.7957 & 0.9160 & 0.8013 & 0.9343 \\
    FedAvg & 0.7960 & 0.9155 & 0.8025 & 0.9351 \\
    FedProx & \underline{0.7964} & 0.9158 & \underline{0.8029} & 0.9351 \\
    Ditto & 0.7962 & \underline{0.9165} & 0.8026 & 0.9345 \\
    FedAMP & 0.7962 & 0.9158 & 0.8027 & \textbf{\underline{0.9352}} \\
    \textbf{PF-MSMTrec} (Fed) & \textbf{0.7965*} & \textbf{0.9202*} & \textbf{0.8083*} & 0.9327 \\
    \bottomrule
    \end{tabular}%
  \label{tab:overall-tenrec-2}%
\end{table}%

\subsubsection{Implementation Details}
We use the same experimental setup for all methods, including the same embedding layers, input features, and training hyper-parameters. 
We employ a three-layer MLP with ReLU activation as the expert network, with hidden layer sizes of \{512, 256, 128\}, and we use a three-layer MLP with sigmoid activation as the tower network, with hidden layer sizes of \{128, 64, 32\}. 
We train all the methods with the binary cross-entropy loss. All methods are optimized using the Adam Optimizer~\cite{adam}. 
The learning rate is set to $0.001$ and the dropout rate is set to $0.2$. For our proposed PF-MSMTrec, the conflict-coordinate hyper-parameter $c$ is set to 0.4 and the coefficient $\lambda$ in the loss function is set to 0.5. 

\begin{table*}[htbp]
\small
  \caption{Ablation study on the AliExpress dataset. Bold text indicates the best performance.}
  \centering
    \begin{tabular}{@{}ccccccccc@{}}
    \toprule
    \textbf{AliExpress} & \multicolumn{2}{c}{\textbf{NL}} & \multicolumn{2}{c}{\textbf{ES}} & \multicolumn{2}{c}{\textbf{FR}} & \multicolumn{2}{c@{}}{\textbf{US}} \\
    \cmidrule(lr){2-3} \cmidrule(lr){4-5} \cmidrule(lr){6-7} \cmidrule(lr){8-9}
     & auc\_ctr & auc\_ctcvr & auc\_ctr & auc\_ctcvr & auc\_ctr & auc\_ctcvr & auc\_ctr & auc\_ctcvr \\
    \midrule
    \textbf{A1} & 0.7274 & 0.8617 & 0.7299 & 0.8902 & 0.7298 & 0.8849 & 0.7081 & 0.8799 \\
    \textbf{A2} & 0.7289 & 0.8511 & 0.7290 & 0.8911 & 0.7305 & 0.8849 & 0.7128 & \textbf{0.8792} \\
    \textbf{A3} & \textbf{0.7316} & \textbf{0.8653} & \textbf{0.7325} & \textbf{0.8925} & \textbf{0.7321} & 0.8851 & \textbf{0.7142} & 0.8791 \\
    \textbf{A4} & 0.7268 & 0.8651 & 0.7314 & 0.8891 & 0.7296 & \textbf{0.8890} & 0.7135 & 0.8740 \\
    \bottomrule
    \end{tabular}%
  \label{tab:abs-ali-1}%
\end{table*}%

\begin{table*}[htbp]
\small
  \caption{Impact of the number of experts on the AliExpress dataset. Bold text indicates the best performance.}
  \centering
    \begin{tabular}{@{}lcccccccc@{}}
    \toprule
    \textbf{AliExpress} & \multicolumn{2}{c}{\textbf{NL}} & \multicolumn{2}{c}{\textbf{FR}} & \multicolumn{2}{c}{\textbf{ES}} & \multicolumn{2}{c@{}}{\textbf{US}} \\
    \cmidrule(lr){2-3} \cmidrule(lr){4-5} \cmidrule(lr){6-7} \cmidrule(lr){8-9}
     & auc\_ctr & auc\_ctcvr & auc\_ctr & auc\_ctcvr & auc\_ctr & auc\_ctcvr & auc\_ctr & auc\_ctcvr \\
    \midrule
    Expert =2 & 0.7288 & 0.8625 & 0.7297 & 0.8892 & 0.7297 & 0.8831 & 0.7127 & 0.8767 \\
    Expert =3 & 0.7313 & 0.8638 & 0.7316 & 0.8915 & 0.7309 & 0.8842 & 0.7130 & 0.8771 \\
    Expert =4 & \textbf{0.7316} & \textbf{0.8653} & \textbf{0.7325} & \textbf{0.8925} & \textbf{0.7321} & \textbf{0.8851} & \textbf{0.7142} & \textbf{0.8791} \\
    Expert =5 & 0.7314 & 0.8644 & 0.7300 & 0.8903 & 0.7316 & 0.8810 & 0.7140 & 0.8185 \\
    Expert =6 & 0.7311 & 0.8649 & 0.7313 & 0.8901 & 0.7307 & 0.8824 & 0.7139 & 0.8347 \\
    \bottomrule
    \end{tabular}%
  \label{tab:abs-ali-2}%
\end{table*}%

In the first group of multi-scenario and multi-task recommendation baseline models, our implementation is as follows: The single-task model is implemented with two separate expert networks and two separate tower networks. For MMoE and PLE, we implement them with four experts, two gate networks, and two towers. MMoE shares all four experts, while PLE shares two of them. For ESMM, we multiply the outputs of the two towers to obtain the CTCVR result. For AITM, we implement the AIT module with three MLPs serving as the query, key, and value of the attention mechanism. Task dependencies are passed through these AIT modules. For STAR, the basic FCN is implemented with center parameters and task-specific parameters. During prediction, the center weight is multiplied by the task-specific weight to generate the output. Additionally, we apply a three-layer MLP as an auxiliary network after shared embedding layers to provide auxiliary information. For AESM$^2$, we employ four experts and select experts using the Kullback-Leibler (KL) divergence. For PEPNet, we replace the gate network with one linear layer and softmax activation by one scale factor and two neural layers with sigmoid and ReLU activation. Additionally, we imitate EPNet by applying an extra element-wise attention network to learn the importance of dimensions in the input embedding.
In the second group of federated learning baseline models, to ensure a fair comparison, we varied only the federated parameter aggregation method when comparing our approach with the baseline methods, while maintaining the local expert network model structure unchanged. For FedProx, the proximal term constant $\mu$ is set to 0.01. For Ditto, the coefficient $\lambda$ that controls the interpolation between the local and global model is set to 0.1. For FedAMP, the hyper-parameter $\alpha_k$ is set to 1.0. 

\begin{table}[t]
\small
  \caption{Ablation study on the Tenrec dataset. Bold text indicates the best performance.}
  \centering
    \begin{tabular}{@{}ccccccccc@{}}
    \toprule
    \textbf{Tenrec} & \multicolumn{2}{c}{\textbf{QK-video}} & \multicolumn{2}{c}{\textbf{QB-article}} \\
    \cmidrule(lr){2-3} \cmidrule(lr){4-5}
     & auc\_click & auc\_like & auc\_click  & auc\_like \\
    \midrule
    \textbf{A1} & 0.7944 & 0.9170 & 0.8066 & 0.9301 \\
    \textbf{A2} & 0.7939 & 0.9192 & 0.8062 & 0.9310 \\
    \textbf{A3} & \textbf{0.7965} & \textbf{0.9202} & \textbf{0.8083} & \textbf{0.9327} \\
    \textbf{A4} & 0.7963 & 0.9195 & 0.8071 & 0.9318 \\
    \bottomrule
    \end{tabular}%
  \label{tab:abs-ten-1}%
\end{table}%

\begin{table}[t]
\small
  \caption{Impact of the number of experts on the Tenrec dataset. Bold text indicates the best performance.}
  \centering
    \begin{tabular}{@{}lcccccccc@{}}
    \toprule
    \textbf{Expert} & \multicolumn{2}{c}{\textbf{QK-video}} & \multicolumn{2}{c}{\textbf{QB-article}} \\
    \cmidrule(lr){2-3} \cmidrule(lr){4-5}
     & auc\_click & auc\_like & auc\_click & auc\_like \\
    \midrule
    Expert =2 & 0.7938 & 0.9178 & 0.8055 & 0.9317 \\
    Expert =3 & 0.7950 & 0.9194 & 0.8071 & 0.9318 \\
    Expert =4 & \textbf{0.7965} & 0.9202 &  \textbf{0.8083} &  \textbf{0.9327} \\
    Expert =5 & 0.7961 & \textbf{0.9204} & 0.8081 & 0.9317 \\
    Expert =6 & 0.7942 & 0.9195 & 0.8077 & 0.9320 \\
    \bottomrule
    \end{tabular}%
  \label{tab:abs-ten-2}%
\end{table}%

\subsubsection{Overall Performance}
Table~\ref{tab:overall-ali-1} and Table~\ref{tab:oveall-ali-2} compare the performance of our method with two groups of baseline models on the AliExpress dataset. Table~\ref{tab:overall-tenrec-1} and Table~\ref{tab:overall-tenrec-2} compare their performance on the Tenrec dataset. 
We have the following observations: (1) Our proposed PF-MSMTrec is evaluated in both federated and non-federated (local) settings. In the non-federated setting, the federated learning module is not required, and all expert networks and tower networks jointly perform predictions. Notably, our proposed method under federated settings even outperforms SOTA multi-scenario multi-task methods under non-federated settings. This indicates that our approach effectively mitigates multiple optimization conflicts and supports joint learning across multiple clients while preserving data privacy. (2) Our method also surpasses SOTA federated learning approaches, demonstrating that our designed federated learning paradigm excels in personalized federated parameter aggregation.


\begin{figure}[t!]
    \centering
    \begin{minipage}{0.23\textwidth}
        \centering
        \includegraphics[width=\linewidth]{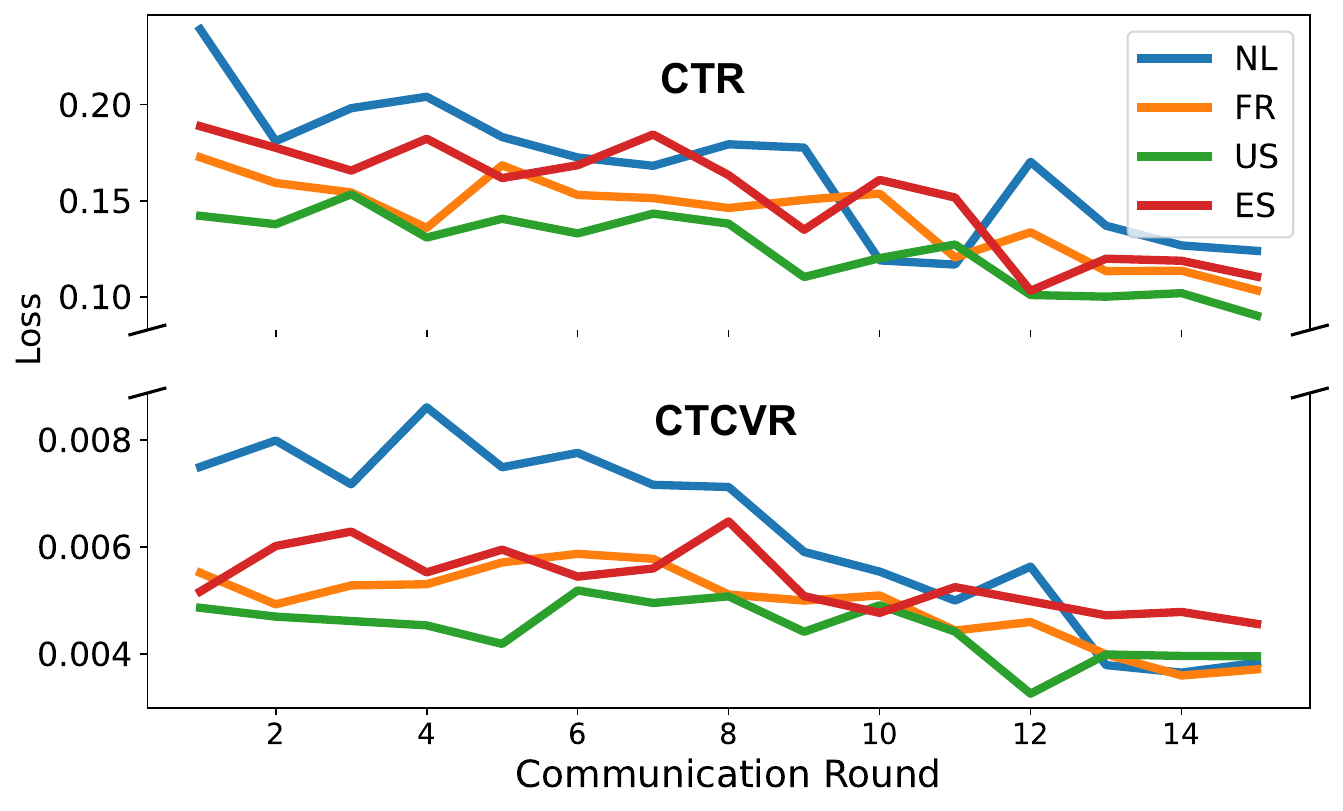}
        \caption*{(a) Ours on AliExpress.}
    \end{minipage}\hfill
    \begin{minipage}{0.23\textwidth}
        \centering
        \includegraphics[width=\linewidth]{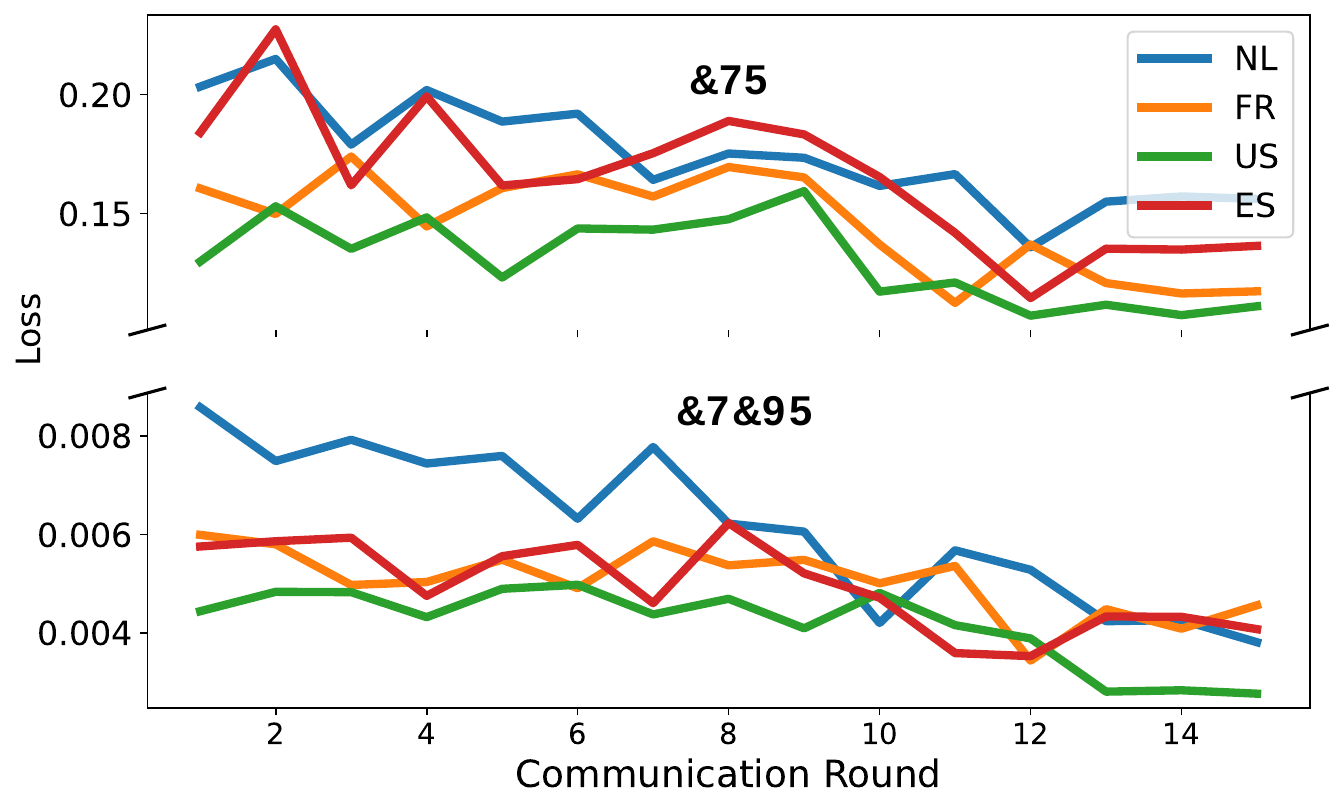}
        \caption*{(b) FedAVG on AliExpress.}
    \end{minipage}\hfill
    \begin{minipage}{0.23\textwidth}
        \centering
        \includegraphics[width=\linewidth]{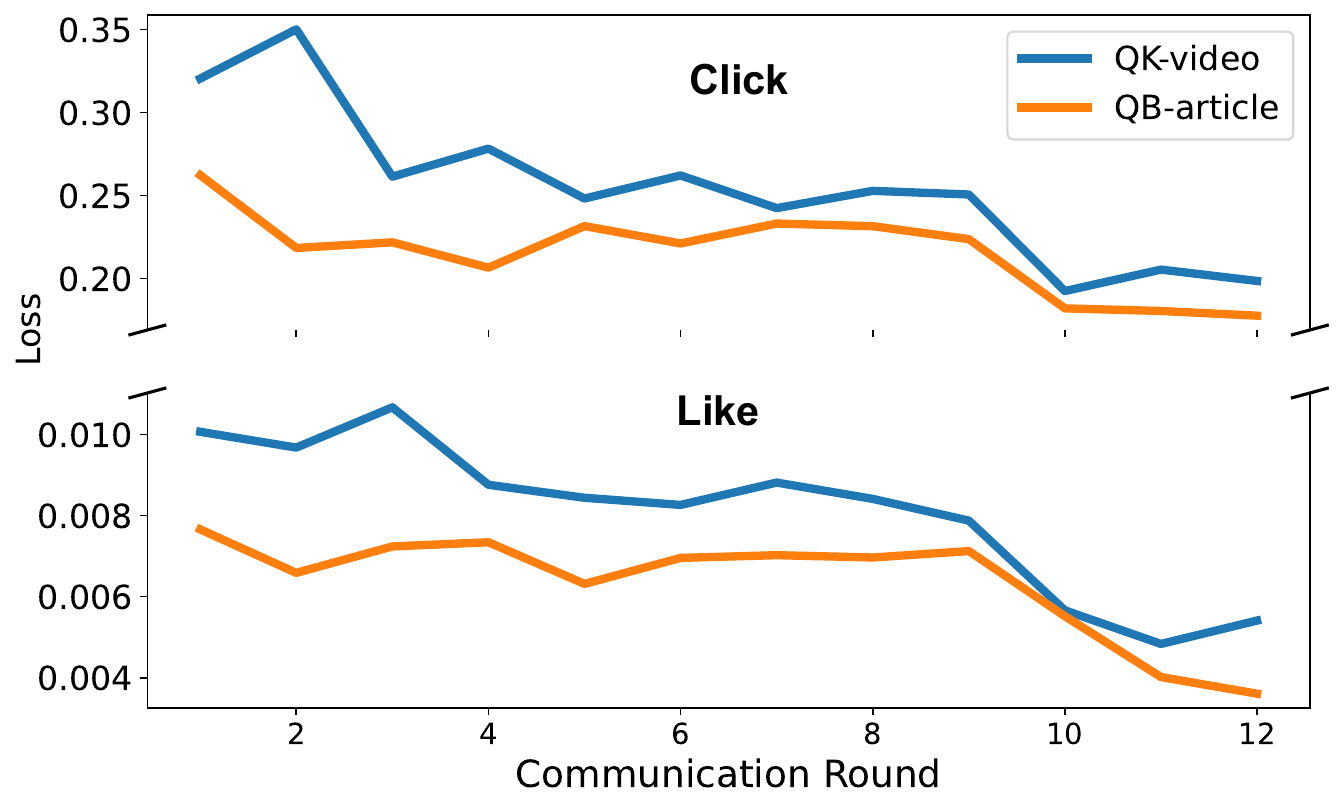}
        \caption*{(c) Ours on Tenrec.}
    \end{minipage}\hfill
    \begin{minipage}{0.23\textwidth}
        \centering
        \includegraphics[width=\linewidth]{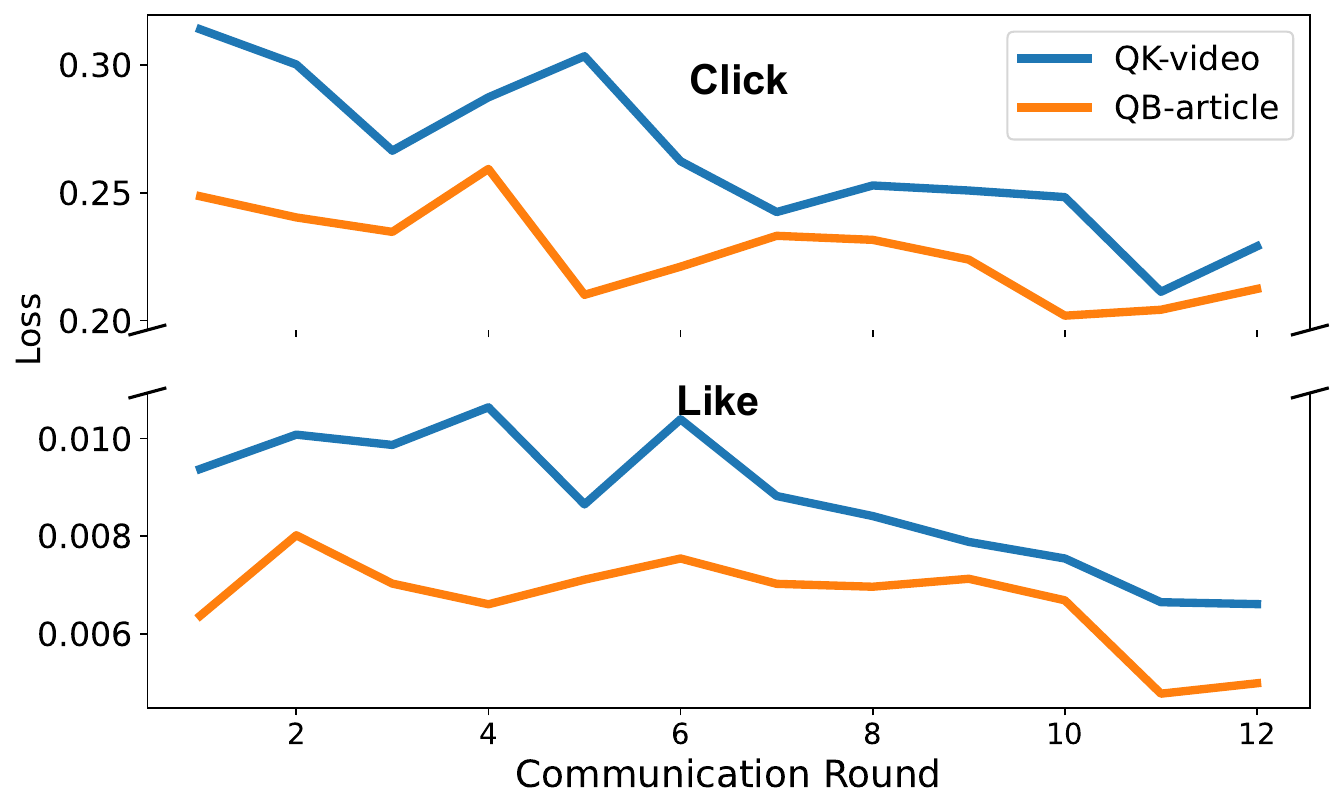}
        \caption*{(d) FedAVG on Tenrec.}
    \end{minipage}
    \caption{Convergence on two datasets.}
    \label{convergence}
\end{figure}

\subsection{In-depth Analysis}
\subsubsection{Ablation study.} We conduct four sets of experiments to investigate the effects of different modules in our model: (\textbf{A1}) Change the aggregation method for both expert and tower networks to FedAvg. (\textbf{A2}) Apply federated averaging to all parameters of the expert network, while keeping the tower network unchanged. (\textbf{A3}) Apply federated averaging only to the scenario-specific parameters of the expert network, while keeping the tower network unchanged. (\textbf{A4}) Apply federated averaging to the tower network, while keeping the expert network unchanged. Table~\ref{tab:abs-ali-1} and Table~\ref{tab:abs-ten-1} describe the results. 
It can be concluded that expert parameter decoupling and conflict coordination in personalized aggregation are crucial for achieving optimal performance.
In contrast, the aggregation method for the tower network has a relatively minor impact.

\subsubsection{The impact of the number of experts.} We test the performance with varying numbers of experts per client, ranging from 2 to 6. Table~\ref{tab:abs-ali-2} and Table~\ref{tab:abs-ten-2} describe the results.
The optimal number of experts is 4, it shows that having too few experts hinders feature extraction, while an excessive number can result in conflicts.

\subsubsection{Convergence study.} 
We compare our method with FedAvg as a baseline and analyze convergence trends on the test dataset. As shown in Figure~\ref{convergence}, our model demonstrates similar convergence behavior to FedAvg and effectively learns multiple tasks across various scenarios on the setting of federated learning.


\section{Conclusion}
In this paper, we explore a new and challenging problem: federated multi-scenario multi-task recommendation. We propose a novel framework called PF-MSMTrec. Our model incorporates parameter decoupling, federated batch normalization, conflict coordination, and personalized aggregation modules. Our proposed method effectively mitigates the multiple optimization conflict issues that arise in such complex application settings. Extensive experiments demonstrate that our proposed model outperforms SOTA methods. We believe that our proposed method broadens the applicability of recommender systems by tackling sophisticated business settings with a federated learning approach.

\balance

\bibliographystyle{ACM-Reference-Format}
\bibliography{main}


\begin{thebibliography}{52}


\ifx \showCODEN    \undefined \def \showCODEN     #1{\unskip}     \fi
\ifx \showDOI      \undefined \def \showDOI       #1{#1}\fi
\ifx \showISBNx    \undefined \def \showISBNx     #1{\unskip}     \fi
\ifx \showISBNxiii \undefined \def \showISBNxiii  #1{\unskip}     \fi
\ifx \showISSN     \undefined \def \showISSN      #1{\unskip}     \fi
\ifx \showLCCN     \undefined \def \showLCCN      #1{\unskip}     \fi
\ifx \shownote     \undefined \def \shownote      #1{#1}          \fi
\ifx \showarticletitle \undefined \def \showarticletitle #1{#1}   \fi
\ifx \showURL      \undefined \def \showURL       {\relax}        \fi
\providecommand\bibfield[2]{#2}
\providecommand\bibinfo[2]{#2}
\providecommand\natexlab[1]{#1}
\providecommand\showeprint[2][]{arXiv:#2}

\bibitem[Agrawal et~al\mbox{.}(2024)]%
        {agrawal2024no}
\bibfield{author}{\bibinfo{person}{Nimesh Agrawal}, \bibinfo{person}{Anuj~Kumar
  Sirohi}, \bibinfo{person}{Sandeep Kumar}, {et~al\mbox{.}}}
  \bibinfo{year}{2024}\natexlab{}.
\newblock \showarticletitle{No Prejudice! Fair Federated Graph Neural Networks
  for Personalized Recommendation}. In \bibinfo{booktitle}{\emph{Proceedings of
  the AAAI Conference on Artificial Intelligence}}, Vol.~\bibinfo{volume}{38}.
  \bibinfo{pages}{10775--10783}.
\newblock


\bibitem[Ammad-Ud-Din et~al\mbox{.}(2019)]%
        {ammad2019federated}
\bibfield{author}{\bibinfo{person}{Muhammad Ammad-Ud-Din},
  \bibinfo{person}{Elena Ivannikova}, \bibinfo{person}{Suleiman~A Khan},
  \bibinfo{person}{Were Oyomno}, \bibinfo{person}{Qiang Fu},
  \bibinfo{person}{Kuan~Eeik Tan}, {and} \bibinfo{person}{Adrian Flanagan}.}
  \bibinfo{year}{2019}\natexlab{}.
\newblock \showarticletitle{Federated collaborative filtering for
  privacy-preserving personalized recommendation system}.
\newblock \bibinfo{journal}{\emph{arXiv preprint arXiv:1901.09888}}
  (\bibinfo{year}{2019}).
\newblock


\bibitem[Bai et~al\mbox{.}(2022)]%
        {bai2022contrastive}
\bibfield{author}{\bibinfo{person}{Ting Bai}, \bibinfo{person}{Yudong Xiao},
  \bibinfo{person}{Bin Wu}, \bibinfo{person}{Guojun Yang},
  \bibinfo{person}{Hongyong Yu}, {and} \bibinfo{person}{Jian-Yun Nie}.}
  \bibinfo{year}{2022}\natexlab{}.
\newblock \showarticletitle{A Contrastive Sharing Model for Multi-Task
  Recommendation}. In \bibinfo{booktitle}{\emph{Proceedings of the ACM Web
  Conference 2022}}. \bibinfo{pages}{3239--3247}.
\newblock


\bibitem[Cai et~al\mbox{.}(2023)]%
        {cai2023many}
\bibfield{author}{\bibinfo{person}{Ruisi Cai}, \bibinfo{person}{Xiaohan Chen},
  \bibinfo{person}{Shiwei Liu}, \bibinfo{person}{Jayanth Srinivasa},
  \bibinfo{person}{Myungjin Lee}, \bibinfo{person}{Ramana Kompella}, {and}
  \bibinfo{person}{Zhangyang Wang}.} \bibinfo{year}{2023}\natexlab{}.
\newblock \showarticletitle{Many-Task Federated Learning: A New Problem Setting
  and a Simple Baseline}. In \bibinfo{booktitle}{\emph{Proceedings of the
  IEEE/CVF Conference on Computer Vision and Pattern Recognition}}.
  \bibinfo{pages}{5036--5044}.
\newblock


\bibitem[Chang et~al\mbox{.}(2023)]%
        {chang2023pepnet}
\bibfield{author}{\bibinfo{person}{Jianxin Chang}, \bibinfo{person}{Chenbin
  Zhang}, \bibinfo{person}{Yiqun Hui}, \bibinfo{person}{Dewei Leng},
  \bibinfo{person}{Yanan Niu}, \bibinfo{person}{Yang Song}, {and}
  \bibinfo{person}{Kun Gai}.} \bibinfo{year}{2023}\natexlab{}.
\newblock \showarticletitle{Pepnet: Parameter and embedding personalized
  network for infusing with personalized prior information}. In
  \bibinfo{booktitle}{\emph{Proceedings of the 29th ACM SIGKDD Conference on
  Knowledge Discovery and Data Mining}}. \bibinfo{pages}{3795--3804}.
\newblock


\bibitem[Chen et~al\mbox{.}(2023)]%
        {chen2023fedbone}
\bibfield{author}{\bibinfo{person}{Yiqiang Chen}, \bibinfo{person}{Teng Zhang},
  \bibinfo{person}{Xinlong Jiang}, \bibinfo{person}{Qian Chen},
  \bibinfo{person}{Chenlong Gao}, {and} \bibinfo{person}{Wuliang Huang}.}
  \bibinfo{year}{2023}\natexlab{}.
\newblock \showarticletitle{Fedbone: Towards large-scale federated multi-task
  learning}.
\newblock \bibinfo{journal}{\emph{arXiv preprint arXiv:2306.17465}}
  (\bibinfo{year}{2023}).
\newblock


\bibitem[Han et~al\mbox{.}(2021)]%
        {han2021deeprec}
\bibfield{author}{\bibinfo{person}{Jialiang Han}, \bibinfo{person}{Yun Ma},
  \bibinfo{person}{Qiaozhu Mei}, {and} \bibinfo{person}{Xuanzhe Liu}.}
  \bibinfo{year}{2021}\natexlab{}.
\newblock \showarticletitle{Deeprec: On-device deep learning for
  privacy-preserving sequential recommendation in mobile commerce}. In
  \bibinfo{booktitle}{\emph{Proceedings of the Web Conference 2021}}.
  \bibinfo{pages}{900--911}.
\newblock


\bibitem[Hao et~al\mbox{.}(2021)]%
        {hao2021adversarial}
\bibfield{author}{\bibinfo{person}{Xiaobo Hao}, \bibinfo{person}{Yudan Liu},
  \bibinfo{person}{Ruobing Xie}, \bibinfo{person}{Kaikai Ge},
  \bibinfo{person}{Linyao Tang}, \bibinfo{person}{Xu Zhang}, {and}
  \bibinfo{person}{Leyu Lin}.} \bibinfo{year}{2021}\natexlab{}.
\newblock \showarticletitle{Adversarial feature translation for multi-domain
  recommendation}. In \bibinfo{booktitle}{\emph{Proceedings of the 27th ACM
  SIGKDD Conference on Knowledge Discovery \& Data Mining}}.
  \bibinfo{pages}{2964--2973}.
\newblock


\bibitem[He et~al\mbox{.}(2022a)]%
        {he2022spreadgnn}
\bibfield{author}{\bibinfo{person}{Chaoyang He}, \bibinfo{person}{Emir Ceyani},
  \bibinfo{person}{Keshav Balasubramanian}, \bibinfo{person}{Murali Annavaram},
  {and} \bibinfo{person}{Salman Avestimehr}.} \bibinfo{year}{2022}\natexlab{a}.
\newblock \showarticletitle{Spreadgnn: Decentralized multi-task federated
  learning for graph neural networks on molecular data}. In
  \bibinfo{booktitle}{\emph{Proceedings of the AAAI Conference on Artificial
  Intelligence}}, Vol.~\bibinfo{volume}{36}. \bibinfo{pages}{6865--6873}.
\newblock


\bibitem[He et~al\mbox{.}(2022b)]%
        {he2022metabalance}
\bibfield{author}{\bibinfo{person}{Yun He}, \bibinfo{person}{Xue Feng},
  \bibinfo{person}{Cheng Cheng}, \bibinfo{person}{Geng Ji},
  \bibinfo{person}{Yunsong Guo}, {and} \bibinfo{person}{James Caverlee}.}
  \bibinfo{year}{2022}\natexlab{b}.
\newblock \showarticletitle{Metabalance: improving multi-task recommendations
  via adapting gradient magnitudes of auxiliary tasks}. In
  \bibinfo{booktitle}{\emph{Proceedings of the ACM Web Conference 2022}}.
  \bibinfo{pages}{2205--2215}.
\newblock


\bibitem[Huan et~al\mbox{.}(2023)]%
        {huan2023samd}
\bibfield{author}{\bibinfo{person}{Zhaoxin Huan}, \bibinfo{person}{Ang Li},
  \bibinfo{person}{Xiaolu Zhang}, \bibinfo{person}{Xu Min},
  \bibinfo{person}{Jieyu Yang}, \bibinfo{person}{Yong He}, {and}
  \bibinfo{person}{Jun Zhou}.} \bibinfo{year}{2023}\natexlab{}.
\newblock \showarticletitle{SAMD: An Industrial Framework for Heterogeneous
  Multi-Scenario Recommendation}. In \bibinfo{booktitle}{\emph{Proceedings of
  the 29th ACM SIGKDD Conference on Knowledge Discovery and Data Mining}}.
  \bibinfo{pages}{4175--4184}.
\newblock


\bibitem[Huang et~al\mbox{.}(2020)]%
        {huang2020federated}
\bibfield{author}{\bibinfo{person}{Mingkai Huang}, \bibinfo{person}{Hao Li},
  \bibinfo{person}{Bing Bai}, \bibinfo{person}{Chang Wang},
  \bibinfo{person}{Kun Bai}, {and} \bibinfo{person}{Fei Wang}.}
  \bibinfo{year}{2020}\natexlab{}.
\newblock \showarticletitle{A federated multi-view deep learning framework for
  privacy-preserving recommendations}.
\newblock \bibinfo{journal}{\emph{arXiv preprint arXiv:2008.10808}}
  (\bibinfo{year}{2020}).
\newblock


\bibitem[Huang et~al\mbox{.}(2021)]%
        {fedamp}
\bibfield{author}{\bibinfo{person}{Yutao Huang}, \bibinfo{person}{Lingyang
  Chu}, \bibinfo{person}{Zirui Zhou}, \bibinfo{person}{Lanjun Wang},
  \bibinfo{person}{Jiangchuan Liu}, \bibinfo{person}{Jian Pei}, {and}
  \bibinfo{person}{Yong Zhang}.} \bibinfo{year}{2021}\natexlab{}.
\newblock \showarticletitle{Personalized cross-silo federated learning on
  non-iid data}, Vol.~\bibinfo{volume}{35}. \bibinfo{pages}{7865--7873}.
\newblock


\bibitem[Jiang et~al\mbox{.}(2022)]%
        {jiang2022adaptive}
\bibfield{author}{\bibinfo{person}{Yuchen Jiang}, \bibinfo{person}{Qi Li},
  \bibinfo{person}{Han Zhu}, \bibinfo{person}{Jinbei Yu}, \bibinfo{person}{Jin
  Li}, \bibinfo{person}{Ziru Xu}, \bibinfo{person}{Huihui Dong}, {and}
  \bibinfo{person}{Bo Zheng}.} \bibinfo{year}{2022}\natexlab{}.
\newblock \showarticletitle{Adaptive domain interest network for multi-domain
  recommendation}. In \bibinfo{booktitle}{\emph{Proceedings of the 31st ACM
  International Conference on Information \& Knowledge Management}}.
  \bibinfo{pages}{3212--3221}.
\newblock


\bibitem[Kingma and Ba(2015)]%
        {adam}
\bibfield{author}{\bibinfo{person}{Diederik~P. Kingma} {and}
  \bibinfo{person}{Jimmy Ba}.} \bibinfo{year}{2015}\natexlab{}.
\newblock \showarticletitle{Adam: {A} Method for Stochastic Optimization}. In
  \bibinfo{booktitle}{\emph{{ICLR} 2015}}.
\newblock


\bibitem[Kone{\v{c}}n{\`y} et~al\mbox{.}(2016)]%
        {konevcny2016federated}
\bibfield{author}{\bibinfo{person}{Jakub Kone{\v{c}}n{\`y}},
  \bibinfo{person}{H~Brendan McMahan}, \bibinfo{person}{Felix~X Yu},
  \bibinfo{person}{Peter Richt{\'a}rik}, \bibinfo{person}{Ananda~Theertha
  Suresh}, {and} \bibinfo{person}{Dave Bacon}.}
  \bibinfo{year}{2016}\natexlab{}.
\newblock \showarticletitle{Federated learning: Strategies for improving
  communication efficiency}.
\newblock \bibinfo{journal}{\emph{arXiv preprint arXiv:1610.05492}}
  (\bibinfo{year}{2016}).
\newblock


\bibitem[Lan et~al\mbox{.}(2023)]%
        {lan2023m3rec}
\bibfield{author}{\bibinfo{person}{Zerong Lan}, \bibinfo{person}{Yingyi Zhang},
  {and} \bibinfo{person}{Xianneng Li}.} \bibinfo{year}{2023}\natexlab{}.
\newblock \showarticletitle{M3REC: A Meta-based Multi-scenario Multi-task
  Recommendation Framework}. In \bibinfo{booktitle}{\emph{Proceedings of the
  17th ACM Conference on Recommender Systems}}. \bibinfo{pages}{771--776}.
\newblock


\bibitem[Li et~al\mbox{.}(2023b)]%
        {li2023adatt}
\bibfield{author}{\bibinfo{person}{Danwei Li}, \bibinfo{person}{Zhengyu Zhang},
  \bibinfo{person}{Siyang Yuan}, \bibinfo{person}{Mingze Gao},
  \bibinfo{person}{Weilin Zhang}, \bibinfo{person}{Chaofei Yang},
  \bibinfo{person}{Xi Liu}, {and} \bibinfo{person}{Jiyan Yang}.}
  \bibinfo{year}{2023}\natexlab{b}.
\newblock \showarticletitle{AdaTT: Adaptive Task-to-Task Fusion Network for
  Multitask Learning in Recommendations}.
\newblock \bibinfo{journal}{\emph{arXiv preprint arXiv:2304.04959}}
  (\bibinfo{year}{2023}).
\newblock


\bibitem[Li et~al\mbox{.}(2021)]%
        {ditto}
\bibfield{author}{\bibinfo{person}{Tian Li}, \bibinfo{person}{Shengyuan Hu},
  \bibinfo{person}{Ahmad Beirami}, {and} \bibinfo{person}{Virginia Smith}.}
  \bibinfo{year}{2021}\natexlab{}.
\newblock \showarticletitle{Ditto: Fair and robust federated learning through
  personalization}. In \bibinfo{booktitle}{\emph{ICML}}.
  \bibinfo{pages}{6357--6368}.
\newblock


\bibitem[Li et~al\mbox{.}(2020)]%
        {fedprox}
\bibfield{author}{\bibinfo{person}{Tian Li}, \bibinfo{person}{Anit~Kumar Sahu},
  \bibinfo{person}{Manzil Zaheer}, \bibinfo{person}{Maziar Sanjabi},
  \bibinfo{person}{Ameet Talwalkar}, {and} \bibinfo{person}{Virginia Smith}.}
  \bibinfo{year}{2020}\natexlab{}.
\newblock \showarticletitle{Federated Optimization in Heterogeneous Networks}.
  In \bibinfo{booktitle}{\emph{MLSys}}.
\newblock


\bibitem[Li et~al\mbox{.}(2023a)]%
        {li2023hamur}
\bibfield{author}{\bibinfo{person}{Xiaopeng Li}, \bibinfo{person}{Fan Yan},
  \bibinfo{person}{Xiangyu Zhao}, \bibinfo{person}{Yichao Wang},
  \bibinfo{person}{Bo Chen}, \bibinfo{person}{Huifeng Guo}, {and}
  \bibinfo{person}{Ruiming Tang}.} \bibinfo{year}{2023}\natexlab{a}.
\newblock \showarticletitle{Hamur: Hyper adapter for multi-domain
  recommendation}. In \bibinfo{booktitle}{\emph{Proceedings of the 32nd ACM
  International Conference on Information and Knowledge Management}}.
  \bibinfo{pages}{1268--1277}.
\newblock


\bibitem[Lin et~al\mbox{.}(2020)]%
        {lin2020fedrec}
\bibfield{author}{\bibinfo{person}{Guanyu Lin}, \bibinfo{person}{Feng Liang},
  \bibinfo{person}{Weike Pan}, {and} \bibinfo{person}{Zhong Ming}.}
  \bibinfo{year}{2020}\natexlab{}.
\newblock \showarticletitle{Fedrec: Federated recommendation with explicit
  feedback}.
\newblock \bibinfo{journal}{\emph{IEEE Intelligent Systems}}
  \bibinfo{volume}{36}, \bibinfo{number}{5} (\bibinfo{year}{2020}),
  \bibinfo{pages}{21--30}.
\newblock


\bibitem[Liu et~al\mbox{.}(2021)]%
        {liu2021conflict}
\bibfield{author}{\bibinfo{person}{Bo Liu}, \bibinfo{person}{Xingchao Liu},
  \bibinfo{person}{Xiaojie Jin}, \bibinfo{person}{Peter Stone}, {and}
  \bibinfo{person}{Qiang Liu}.} \bibinfo{year}{2021}\natexlab{}.
\newblock \showarticletitle{Conflict-averse gradient descent for multi-task
  learning}.
\newblock \bibinfo{journal}{\emph{Advances in Neural Information Processing
  Systems}}  \bibinfo{volume}{34} (\bibinfo{year}{2021}),
  \bibinfo{pages}{18878--18890}.
\newblock


\bibitem[Luo et~al\mbox{.}(2022)]%
        {luo2022personalized}
\bibfield{author}{\bibinfo{person}{Sichun Luo}, \bibinfo{person}{Yuanzhang
  Xiao}, {and} \bibinfo{person}{Linqi Song}.} \bibinfo{year}{2022}\natexlab{}.
\newblock \showarticletitle{Personalized federated recommendation via joint
  representation learning, user clustering, and model adaptation}. In
  \bibinfo{booktitle}{\emph{Proceedings of the 31st ACM international
  conference on information \& knowledge management}}.
  \bibinfo{pages}{4289--4293}.
\newblock


\bibitem[Ma et~al\mbox{.}(2018b)]%
        {ma2018modeling}
\bibfield{author}{\bibinfo{person}{Jiaqi Ma}, \bibinfo{person}{Zhe Zhao},
  \bibinfo{person}{Xinyang Yi}, \bibinfo{person}{Jilin Chen},
  \bibinfo{person}{Lichan Hong}, {and} \bibinfo{person}{Ed~H Chi}.}
  \bibinfo{year}{2018}\natexlab{b}.
\newblock \showarticletitle{Modeling task relationships in multi-task learning
  with multi-gate mixture-of-experts}. In \bibinfo{booktitle}{\emph{Proceedings
  of the 24th ACM SIGKDD international conference on knowledge discovery \&
  data mining}}. \bibinfo{pages}{1930--1939}.
\newblock


\bibitem[Ma et~al\mbox{.}(2018a)]%
        {ma2018entire}
\bibfield{author}{\bibinfo{person}{Xiao Ma}, \bibinfo{person}{Liqin Zhao},
  \bibinfo{person}{Guan Huang}, \bibinfo{person}{Zhi Wang},
  \bibinfo{person}{Zelin Hu}, \bibinfo{person}{Xiaoqiang Zhu}, {and}
  \bibinfo{person}{Kun Gai}.} \bibinfo{year}{2018}\natexlab{a}.
\newblock \showarticletitle{Entire space multi-task model: An effective
  approach for estimating post-click conversion rate}. In
  \bibinfo{booktitle}{\emph{The 41st International ACM SIGIR Conference on
  Research \& Development in Information Retrieval}}.
  \bibinfo{pages}{1137--1140}.
\newblock


\bibitem[Maeng et~al\mbox{.}(2022)]%
        {maeng2022towards}
\bibfield{author}{\bibinfo{person}{Kiwan Maeng}, \bibinfo{person}{Haiyu Lu},
  \bibinfo{person}{Luca Melis}, \bibinfo{person}{John Nguyen},
  \bibinfo{person}{Mike Rabbat}, {and} \bibinfo{person}{Carole-Jean Wu}.}
  \bibinfo{year}{2022}\natexlab{}.
\newblock \showarticletitle{Towards fair federated recommendation learning:
  Characterizing the inter-dependence of system and data heterogeneity}. In
  \bibinfo{booktitle}{\emph{Proceedings of the 16th ACM Conference on
  Recommender Systems}}. \bibinfo{pages}{156--167}.
\newblock


\bibitem[Marfoq et~al\mbox{.}(2021)]%
        {marfoq2021federated}
\bibfield{author}{\bibinfo{person}{Othmane Marfoq}, \bibinfo{person}{Giovanni
  Neglia}, \bibinfo{person}{Aur{\'e}lien Bellet}, \bibinfo{person}{Laetitia
  Kameni}, {and} \bibinfo{person}{Richard Vidal}.}
  \bibinfo{year}{2021}\natexlab{}.
\newblock \showarticletitle{Federated multi-task learning under a mixture of
  distributions}.
\newblock \bibinfo{journal}{\emph{Advances in Neural Information Processing
  Systems}}  \bibinfo{volume}{34} (\bibinfo{year}{2021}),
  \bibinfo{pages}{15434--15447}.
\newblock


\bibitem[McMahan et~al\mbox{.}(2017)]%
        {mcmahan2017communication}
\bibfield{author}{\bibinfo{person}{Brendan McMahan}, \bibinfo{person}{Eider
  Moore}, \bibinfo{person}{Daniel Ramage}, \bibinfo{person}{Seth Hampson},
  {and} \bibinfo{person}{Blaise~Aguera y Arcas}.}
  \bibinfo{year}{2017}\natexlab{}.
\newblock \showarticletitle{Communication-efficient learning of deep networks
  from decentralized data}. In \bibinfo{booktitle}{\emph{Artificial
  intelligence and statistics}}. PMLR, \bibinfo{pages}{1273--1282}.
\newblock


\bibitem[Mills et~al\mbox{.}(2021)]%
        {mills2021multi}
\bibfield{author}{\bibinfo{person}{Jed Mills}, \bibinfo{person}{Jia Hu}, {and}
  \bibinfo{person}{Geyong Min}.} \bibinfo{year}{2021}\natexlab{}.
\newblock \showarticletitle{Multi-task federated learning for personalised deep
  neural networks in edge computing}.
\newblock \bibinfo{journal}{\emph{IEEE Transactions on Parallel and Distributed
  Systems}} \bibinfo{volume}{33}, \bibinfo{number}{3} (\bibinfo{year}{2021}),
  \bibinfo{pages}{630--641}.
\newblock


\bibitem[Muhammad et~al\mbox{.}(2020)]%
        {muhammad2020fedfast}
\bibfield{author}{\bibinfo{person}{Khalil Muhammad}, \bibinfo{person}{Qinqin
  Wang}, \bibinfo{person}{Diarmuid O'Reilly-Morgan}, \bibinfo{person}{Elias
  Tragos}, \bibinfo{person}{Barry Smyth}, \bibinfo{person}{Neil Hurley},
  \bibinfo{person}{James Geraci}, {and} \bibinfo{person}{Aonghus Lawlor}.}
  \bibinfo{year}{2020}\natexlab{}.
\newblock \showarticletitle{Fedfast: Going beyond average for faster training
  of federated recommender systems}. In \bibinfo{booktitle}{\emph{Proceedings
  of the 26th ACM SIGKDD International Conference on Knowledge Discovery \&
  Data Mining}}. \bibinfo{pages}{1234--1242}.
\newblock


\bibitem[Ning et~al\mbox{.}(2023)]%
        {ning2023multi}
\bibfield{author}{\bibinfo{person}{Wentao Ning}, \bibinfo{person}{Xiao Yan},
  \bibinfo{person}{Weiwen Liu}, \bibinfo{person}{Reynold Cheng},
  \bibinfo{person}{Rui Zhang}, {and} \bibinfo{person}{Bo Tang}.}
  \bibinfo{year}{2023}\natexlab{}.
\newblock \showarticletitle{Multi-domain Recommendation with Embedding
  Disentangling and Domain Alignment}. In \bibinfo{booktitle}{\emph{Proceedings
  of the 32nd ACM International Conference on Information and Knowledge
  Management}}. \bibinfo{pages}{1917--1927}.
\newblock


\bibitem[Qu et~al\mbox{.}(2023)]%
        {qu2023semi}
\bibfield{author}{\bibinfo{person}{Liang Qu}, \bibinfo{person}{Ningzhi Tang},
  \bibinfo{person}{Ruiqi Zheng}, \bibinfo{person}{Quoc Viet~Hung Nguyen},
  \bibinfo{person}{Zi Huang}, \bibinfo{person}{Yuhui Shi}, {and}
  \bibinfo{person}{Hongzhi Yin}.} \bibinfo{year}{2023}\natexlab{}.
\newblock \showarticletitle{Semi-decentralized federated ego graph learning for
  recommendation}. In \bibinfo{booktitle}{\emph{Proceedings of the ACM Web
  Conference 2023}}. \bibinfo{pages}{339--348}.
\newblock


\bibitem[Sheng et~al\mbox{.}(2021)]%
        {sheng2021one}
\bibfield{author}{\bibinfo{person}{Xiang-Rong Sheng}, \bibinfo{person}{Liqin
  Zhao}, \bibinfo{person}{Guorui Zhou}, \bibinfo{person}{Xinyao Ding},
  \bibinfo{person}{Binding Dai}, \bibinfo{person}{Qiang Luo},
  \bibinfo{person}{Siran Yang}, \bibinfo{person}{Jingshan Lv},
  \bibinfo{person}{Chi Zhang}, \bibinfo{person}{Hongbo Deng}, {et~al\mbox{.}}}
  \bibinfo{year}{2021}\natexlab{}.
\newblock \showarticletitle{One model to serve all: Star topology adaptive
  recommender for multi-domain ctr prediction}. In
  \bibinfo{booktitle}{\emph{Proceedings of the 30th ACM International
  Conference on Information \& Knowledge Management}}.
  \bibinfo{pages}{4104--4113}.
\newblock


\bibitem[Smith et~al\mbox{.}(2017)]%
        {smith2017federated}
\bibfield{author}{\bibinfo{person}{Virginia Smith}, \bibinfo{person}{Chao-Kai
  Chiang}, \bibinfo{person}{Maziar Sanjabi}, {and} \bibinfo{person}{Ameet~S
  Talwalkar}.} \bibinfo{year}{2017}\natexlab{}.
\newblock \showarticletitle{Federated multi-task learning}.
\newblock \bibinfo{journal}{\emph{Advances in neural information processing
  systems}}  \bibinfo{volume}{30} (\bibinfo{year}{2017}).
\newblock


\bibitem[Tan et~al\mbox{.}(2022)]%
        {tan2022towards}
\bibfield{author}{\bibinfo{person}{Alysa~Ziying Tan}, \bibinfo{person}{Han Yu},
  \bibinfo{person}{Lizhen Cui}, {and} \bibinfo{person}{Qiang Yang}.}
  \bibinfo{year}{2022}\natexlab{}.
\newblock \showarticletitle{Towards personalized federated learning}.
\newblock \bibinfo{journal}{\emph{IEEE Transactions on Neural Networks and
  Learning Systems}} (\bibinfo{year}{2022}).
\newblock


\bibitem[Tan et~al\mbox{.}(2020)]%
        {tan2020federated}
\bibfield{author}{\bibinfo{person}{Ben Tan}, \bibinfo{person}{Bo Liu},
  \bibinfo{person}{Vincent Zheng}, {and} \bibinfo{person}{Qiang Yang}.}
  \bibinfo{year}{2020}\natexlab{}.
\newblock \showarticletitle{A federated recommender system for online
  services}. In \bibinfo{booktitle}{\emph{Proceedings of the 14th ACM
  Conference on Recommender Systems}}. \bibinfo{pages}{579--581}.
\newblock


\bibitem[Tang et~al\mbox{.}(2020)]%
        {tang2020progressive}
\bibfield{author}{\bibinfo{person}{Hongyan Tang}, \bibinfo{person}{Junning
  Liu}, \bibinfo{person}{Ming Zhao}, {and} \bibinfo{person}{Xudong Gong}.}
  \bibinfo{year}{2020}\natexlab{}.
\newblock \showarticletitle{Progressive layered extraction (ple): A novel
  multi-task learning (mtl) model for personalized recommendations}. In
  \bibinfo{booktitle}{\emph{Proceedings of the 14th ACM Conference on
  Recommender Systems}}. \bibinfo{pages}{269--278}.
\newblock


\bibitem[Tian et~al\mbox{.}(2023)]%
        {tian2023multi}
\bibfield{author}{\bibinfo{person}{Yu Tian}, \bibinfo{person}{Bofang Li},
  \bibinfo{person}{Si Chen}, \bibinfo{person}{Xubin Li},
  \bibinfo{person}{Hongbo Deng}, \bibinfo{person}{Jian Xu}, \bibinfo{person}{Bo
  Zheng}, \bibinfo{person}{Qian Wang}, {and} \bibinfo{person}{Chenliang Li}.}
  \bibinfo{year}{2023}\natexlab{}.
\newblock \showarticletitle{Multi-Scenario Ranking with Adaptive Feature
  Learning}. In \bibinfo{booktitle}{\emph{Proceedings of the 46th International
  ACM SIGIR Conference on Research and Development in Information Retrieval}}.
  \bibinfo{pages}{517--526}.
\newblock


\bibitem[Vandenhende et~al\mbox{.}(2021)]%
        {mtl-cv1}
\bibfield{author}{\bibinfo{person}{Simon Vandenhende},
  \bibinfo{person}{Stamatios Georgoulis}, \bibinfo{person}{Wouter
  Van~Gansbeke}, \bibinfo{person}{Marc Proesmans}, \bibinfo{person}{Dengxin
  Dai}, {and} \bibinfo{person}{Luc Van~Gool}.} \bibinfo{year}{2021}\natexlab{}.
\newblock \showarticletitle{Multi-task learning for dense prediction tasks: A
  survey}.
\newblock \bibinfo{journal}{\emph{IEEE transactions on pattern analysis and
  machine intelligence}} \bibinfo{volume}{44}, \bibinfo{number}{7}
  (\bibinfo{year}{2021}), \bibinfo{pages}{3614--3633}.
\newblock


\bibitem[Wang et~al\mbox{.}(2022)]%
        {wang2022multi}
\bibfield{author}{\bibinfo{person}{Sinan Wang}, \bibinfo{person}{Yumeng Li},
  \bibinfo{person}{Hongyan Li}, \bibinfo{person}{Tanchao Zhu},
  \bibinfo{person}{Zhao Li}, {and} \bibinfo{person}{Wenwu Ou}.}
  \bibinfo{year}{2022}\natexlab{}.
\newblock \showarticletitle{Multi-task learning with calibrated mixture of
  insightful experts}. In \bibinfo{booktitle}{\emph{2022 IEEE 38th
  International Conference on Data Engineering (ICDE)}}. IEEE,
  \bibinfo{pages}{3307--3319}.
\newblock


\bibitem[Wang et~al\mbox{.}(2023a)]%
        {wang2023multi}
\bibfield{author}{\bibinfo{person}{Yuhao Wang}, \bibinfo{person}{Ha~Tsz Lam},
  \bibinfo{person}{Yi Wong}, \bibinfo{person}{Ziru Liu},
  \bibinfo{person}{Xiangyu Zhao}, \bibinfo{person}{Yichao Wang},
  \bibinfo{person}{Bo Chen}, \bibinfo{person}{Huifeng Guo}, {and}
  \bibinfo{person}{Ruiming Tang}.} \bibinfo{year}{2023}\natexlab{a}.
\newblock \showarticletitle{Multi-Task Deep Recommender Systems: A Survey}.
\newblock \bibinfo{journal}{\emph{arXiv preprint arXiv:2302.03525}}
  (\bibinfo{year}{2023}).
\newblock


\bibitem[Wang et~al\mbox{.}(2023b)]%
        {wang2023plate}
\bibfield{author}{\bibinfo{person}{Yuhao Wang}, \bibinfo{person}{Xiangyu Zhao},
  \bibinfo{person}{Bo Chen}, \bibinfo{person}{Qidong Liu},
  \bibinfo{person}{Huifeng Guo}, \bibinfo{person}{Huanshuo Liu},
  \bibinfo{person}{Yichao Wang}, \bibinfo{person}{Rui Zhang}, {and}
  \bibinfo{person}{Ruiming Tang}.} \bibinfo{year}{2023}\natexlab{b}.
\newblock \showarticletitle{PLATE: A Prompt-Enhanced Paradigm for
  Multi-Scenario Recommendations}. In \bibinfo{booktitle}{\emph{Proceedings of
  the 46th International ACM SIGIR Conference on Research and Development in
  Information Retrieval}}. \bibinfo{pages}{1498--1507}.
\newblock


\bibitem[Wu et~al\mbox{.}(2021)]%
        {wu2021fedgnn}
\bibfield{author}{\bibinfo{person}{Chuhan Wu}, \bibinfo{person}{Fangzhao Wu},
  \bibinfo{person}{Yang Cao}, \bibinfo{person}{Yongfeng Huang}, {and}
  \bibinfo{person}{Xing Xie}.} \bibinfo{year}{2021}\natexlab{}.
\newblock \showarticletitle{Fedgnn: Federated graph neural network for
  privacy-preserving recommendation}.
\newblock \bibinfo{journal}{\emph{arXiv preprint arXiv:2102.04925}}
  (\bibinfo{year}{2021}).
\newblock


\bibitem[Xi et~al\mbox{.}(2021)]%
        {xi2021modeling}
\bibfield{author}{\bibinfo{person}{Dongbo Xi}, \bibinfo{person}{Zhen Chen},
  \bibinfo{person}{Peng Yan}, \bibinfo{person}{Yinger Zhang},
  \bibinfo{person}{Yongchun Zhu}, \bibinfo{person}{Fuzhen Zhuang}, {and}
  \bibinfo{person}{Yu Chen}.} \bibinfo{year}{2021}\natexlab{}.
\newblock \showarticletitle{Modeling the sequential dependence among audience
  multi-step conversions with multi-task learning in targeted display
  advertising}. In \bibinfo{booktitle}{\emph{Proceedings of the 27th ACM SIGKDD
  Conference on Knowledge Discovery \& Data Mining}}.
  \bibinfo{pages}{3745--3755}.
\newblock


\bibitem[Yang et~al\mbox{.}(2019)]%
        {yang2019federated}
\bibfield{author}{\bibinfo{person}{Qiang Yang}, \bibinfo{person}{Yang Liu},
  \bibinfo{person}{Tianjian Chen}, {and} \bibinfo{person}{Yongxin Tong}.}
  \bibinfo{year}{2019}\natexlab{}.
\newblock \showarticletitle{Federated machine learning: Concept and
  applications}.
\newblock \bibinfo{journal}{\emph{ACM Transactions on Intelligent Systems and
  Technology (TIST)}} \bibinfo{volume}{10}, \bibinfo{number}{2}
  (\bibinfo{year}{2019}), \bibinfo{pages}{1--19}.
\newblock


\bibitem[Zhang et~al\mbox{.}(2023b)]%
        {zhang2023dual}
\bibfield{author}{\bibinfo{person}{Chunxu Zhang}, \bibinfo{person}{Guodong
  Long}, \bibinfo{person}{Tianyi Zhou}, \bibinfo{person}{Peng Yan},
  \bibinfo{person}{Zijian Zhang}, \bibinfo{person}{Chengqi Zhang}, {and}
  \bibinfo{person}{Bo Yang}.} \bibinfo{year}{2023}\natexlab{b}.
\newblock \showarticletitle{Dual personalization on federated recommendation}.
\newblock \bibinfo{journal}{\emph{arXiv preprint arXiv:2301.08143}}
  (\bibinfo{year}{2023}).
\newblock


\bibitem[Zhang et~al\mbox{.}(2023a)]%
        {zhang2023meta}
\bibfield{author}{\bibinfo{person}{Yingyi Zhang}, \bibinfo{person}{Xianneng
  Li}, \bibinfo{person}{Yahe Yu}, \bibinfo{person}{Jian Tang},
  \bibinfo{person}{Huanfang Deng}, \bibinfo{person}{Junya Lu},
  \bibinfo{person}{Yeyin Zhang}, \bibinfo{person}{Qiancheng Jiang},
  \bibinfo{person}{Yunsen Xian}, \bibinfo{person}{Liqian Yu}, {et~al\mbox{.}}}
  \bibinfo{year}{2023}\natexlab{a}.
\newblock \showarticletitle{Meta-generator enhanced multi-domain
  recommendation}. In \bibinfo{booktitle}{\emph{Companion Proceedings of the
  ACM Web Conference 2023}}. \bibinfo{pages}{485--489}.
\newblock


\bibitem[Zhang et~al\mbox{.}(2023c)]%
        {mtl-nlp1}
\bibfield{author}{\bibinfo{person}{Zhihan Zhang}, \bibinfo{person}{Wenhao Yu},
  \bibinfo{person}{Mengxia Yu}, \bibinfo{person}{Zhichun Guo}, {and}
  \bibinfo{person}{Meng Jiang}.} \bibinfo{year}{2023}\natexlab{c}.
\newblock \showarticletitle{A Survey of Multi-task Learning in Natural Language
  Processing: Regarding Task Relatedness and Training Methods}. In
  \bibinfo{booktitle}{\emph{{EACL}}}. \bibinfo{pages}{943--956}.
\newblock


\bibitem[Zhou et~al\mbox{.}(2023)]%
        {zhou2023hinet}
\bibfield{author}{\bibinfo{person}{Jie Zhou}, \bibinfo{person}{Xianshuai Cao},
  \bibinfo{person}{Wenhao Li}, \bibinfo{person}{Lin Bo}, \bibinfo{person}{Kun
  Zhang}, \bibinfo{person}{Chuan Luo}, {and} \bibinfo{person}{Qian Yu}.}
  \bibinfo{year}{2023}\natexlab{}.
\newblock \showarticletitle{Hinet: Novel multi-scenario \& multi-task learning
  with hierarchical information extraction}. In \bibinfo{booktitle}{\emph{2023
  IEEE 39th International Conference on Data Engineering (ICDE)}}. IEEE,
  \bibinfo{pages}{2969--2975}.
\newblock


\bibitem[Zhuang et~al\mbox{.}(2023)]%
        {zhuang2023mas}
\bibfield{author}{\bibinfo{person}{Weiming Zhuang}, \bibinfo{person}{Yonggang
  Wen}, \bibinfo{person}{Lingjuan Lyu}, {and} \bibinfo{person}{Shuai Zhang}.}
  \bibinfo{year}{2023}\natexlab{}.
\newblock \showarticletitle{MAS: Towards Resource-Efficient Federated
  Multiple-Task Learning}. In \bibinfo{booktitle}{\emph{Proceedings of the
  IEEE/CVF International Conference on Computer Vision}}.
  \bibinfo{pages}{23414--23424}.
\newblock


\bibitem[Zou et~al\mbox{.}(2022)]%
        {zou2022automatic}
\bibfield{author}{\bibinfo{person}{Xinyu Zou}, \bibinfo{person}{Zhi Hu},
  \bibinfo{person}{Yiming Zhao}, \bibinfo{person}{Xuchu Ding},
  \bibinfo{person}{Zhongyi Liu}, \bibinfo{person}{Chenliang Li}, {and}
  \bibinfo{person}{Aixin Sun}.} \bibinfo{year}{2022}\natexlab{}.
\newblock \showarticletitle{Automatic expert selection for multi-scenario and
  multi-task search}. In \bibinfo{booktitle}{\emph{Proceedings of the 45th
  International ACM SIGIR Conference on Research and Development in Information
  Retrieval}}. \bibinfo{pages}{1535--1544}.
\newblock


\end{thebibliography}

\clearpage

\appendix

\section{The Proof in Section 3.3.1}

\begin{minipage}{0.45\textwidth}

To proof that the aggregated parameter $\mathbf{\bar{W}}_j^n$ is only related to the normalization parameter $\beta_g$ if federated communication occurs after the completion of each local batch, we substitute Eq. (8) into Eq. (10), we can derive:

\begin{align*}
\mathbf{\bar{W}}_j^n &= \frac{1}{N \cdot S} \sum_{n=1}^{N} \sum_{j=1}^{S} \text{FedBN}(\mathbf{W}_j^n) \\
&= \frac{1}{N \cdot S} \sum_{n=1}^{N} \sum_{j=1}^{S} \left[ \frac{\gamma_g (\mathbf{W}_j^n - \mathbf{\mu}_g)}{\sqrt{\mathbf{\sigma}_g^2 + \mathbf{\epsilon}_g}} + \mathbf{\beta}_g \right] \quad (\text{Eq}.~{(8)})\\
&= \frac{1}{N \cdot S} \cdot \frac{\gamma_g}{\sqrt{\mathbf{\sigma}_g^2 + \mathbf{\epsilon}_g}} \cdot \sum_{n=1}^{N} \sum_{j=1}^{S} (\mathbf{W}_j^n - \mathbf{\mu}_g) + \mathbf{\beta}_g \\
&= \frac{1}{N \cdot S} \cdot \frac{\gamma_g}{\sqrt{\mathbf{\sigma}_g^2 + \mathbf{\epsilon}_g}} \cdot \left( \sum_{n=1}^{N} \sum_{j=1}^{S} \mathbf{W}_j^n - N \cdot S \cdot \mathbf{\mu}_g \right) + \mathbf{\beta}_g \\
&= \frac{\gamma_g}{\sqrt{\mathbf{\sigma}_g^2 + \mathbf{\epsilon}_g}} (\mathbf{\mu}_g - \mathbf{\mu}_g) + \mathbf{\beta}_g \quad (\text{Eq}.~(7))\\
&= \beta_g
\end{align*}

A communication round typically occurs after one or more local training epochs, with each epoch comprising multiple batches. Consequently, the $\beta$ value at the communication round differs from that at an individual batch.

\end{minipage}

\end{document}